\documentclass[12pt]{iopart}

\usepackage{iopams}
\usepackage{graphicx}
\usepackage[T1]{fontenc}
\usepackage{color}
\usepackage{textcomp}

\begin{document}

\title[Phonon coupling in hBN color centers]{Phonon-assisted emission and absorption of individual\\ color centers in hexagonal boron nitride}

\author{Daniel~Wigger$^{1}$, Robert~Schmidt$^{2}$, Osvaldo Del Pozo-Zamudio$^{2,3}$, Johann~A.~Preu\ss{}$^{2}$, Philipp~Tonndorf$^2$, Robert~Schneider$^2$, Paul~Steeger$^2$, Johannes~Kern$^2$, Yashar Khodaei$^4$, Jaroslaw~Sperling$^4$, Steffen Michaelis de Vasconcellos$^2$, Rudolf~Bratschitsch$^2$ and Tilmann~Kuhn$^1$}

\address{$^1$Institute of Solid State Theory, University of M\"{u}nster, 48149 M\"{u}nster, Germany}
\address{$^2$Institute of Physics and Center for Nanotechnology, University of M\"{u}nster, 48149~M\"{u}nster, Germany}
\address{$^3$Instituto de Investigaci\'on en Comunicaci\'on Optica, Universidad Aut\'onoma de San Luis Potos\'i, 78210 San Luis Potosi, M\'exico}
\address{$^4$Terahertz Technology and Photonics, H\"ubner GmbH \& Co K.G., 34123 Kassel, Germany}

\ead{d.wigger@wwu.de}

\begin{abstract}
Defect centers in hexagonal boron nitride represent room-temperature single-photon sources in a layered van der Waals material. These light emitters appear with a wide range of transition energies ranging over the entire visible spectrum, which renders the identification of the underlying atomic structure challenging. In addition to their eminent properties as quantum light emitters, the coupling to phonons is remarkable. Their photoluminescence exhibits significant side band emission well separated from the zero phonon line (ZPL) and an asymmetric broadening of the ZPL itself. In this combined theoretical and experimental study we show that the phonon side bands can be well described in terms of the coupling to bulk longitudinal optical (LO) phonons. To describe the ZPL asymmetry we show that in addition to the coupling to longitudinal acoustic (LA) phonons also the coupling to local mode oscillations of the defect center with respect to the entire host crystal has to be considered. By studying the influence of the emitter's wave function dimensions on the phonon side bands we find reasonable values for size of the wave function and the deformation potentials. We perform photoluminescence excitation measurements to demonstrate that the excitation of the emitters is most efficient by LO-phonon assisted absorption.
\end{abstract}

\maketitle

\section{Introduction}

Single-photon emitters are at the heart of many promising quantum technologies such as quantum computing and quantum cryptography. Although there are various solid-state emitters of single photons, there is still no system available, which simultaneously meets all requirements~\cite{aharonovich2016sol}. For example, self-assembled semiconductor quantum dots~\cite{santori2001tri, michler2003sin} represent a mature technology platform, but presently yield optimum performance only at cryogenic temperatures. Another class of single-photon emitters are defect centers in insulators, such as color centers in diamond~\cite{zaitsev2000vib, jelezko2006sin, beha2012opt}. These defect centers exhibit prominent single-photon emission also at room temperature but suffer from a high refractive index and challenging processability of the host crystal. Recently, a new class of single-photon emitters in atomically thin semiconductors has been discovered~\cite{tonndorf2015sin, koperski2015sin, srivastava2015opt, he2015sin, chakraborty2015vol, tonndorf2017sin, palacios2016ato} and has been deterministically positioned on the nanoscale by strain engineering~\cite{kern2016nan, branny2017det, palacios2017lar}. Defect centers in hexagonal boron-nitride (hBN) combine features from these classes~\cite{tran2016quaI, tran2016quaII,tran2016rob,vuong2016pho,martinez2016eff,jungwirth2016temp,shotan2016pho,exarhos2017opt,li2017non,jungwirth2017opt,sontheimer2017pho,grosso2017tun,tran2017res,chejanovsky2017qua,noh2018sta,xu2018sin}. They have the characteristics of an atomically sized defect center and at the same time share the advantages of a layered structure, i.e., the ultimate limit of miniaturization due to their atomic thickness and high mechanical robustness.

In this work, we investigate the photoluminescence (PL) and photon absorption of defect centers in hBN nanocrystals. By developing a theoretical model to calculate the PL spectrum taking into account the coupling to bulk longitudinal optical (LO) and longitudinal acoustic (LA) phonons and to the oscillation of a local mode, we explain measured emission spectra achieving an excellent agreement between experiment and theory. In addition, we verify by means of photoluminescence excitation (PLE) spectroscopy that coupling to LO phonons provides an efficient way of exciting individual single-photon emitters. This then allows for the isolation of a desired single-photon emitter from the wide range of emission energies of various defect centers in hBN.

\section{Results}
\subsection{Emission spectra}

\begin{figure}[h]
\centering
\includegraphics[width=0.55\columnwidth]{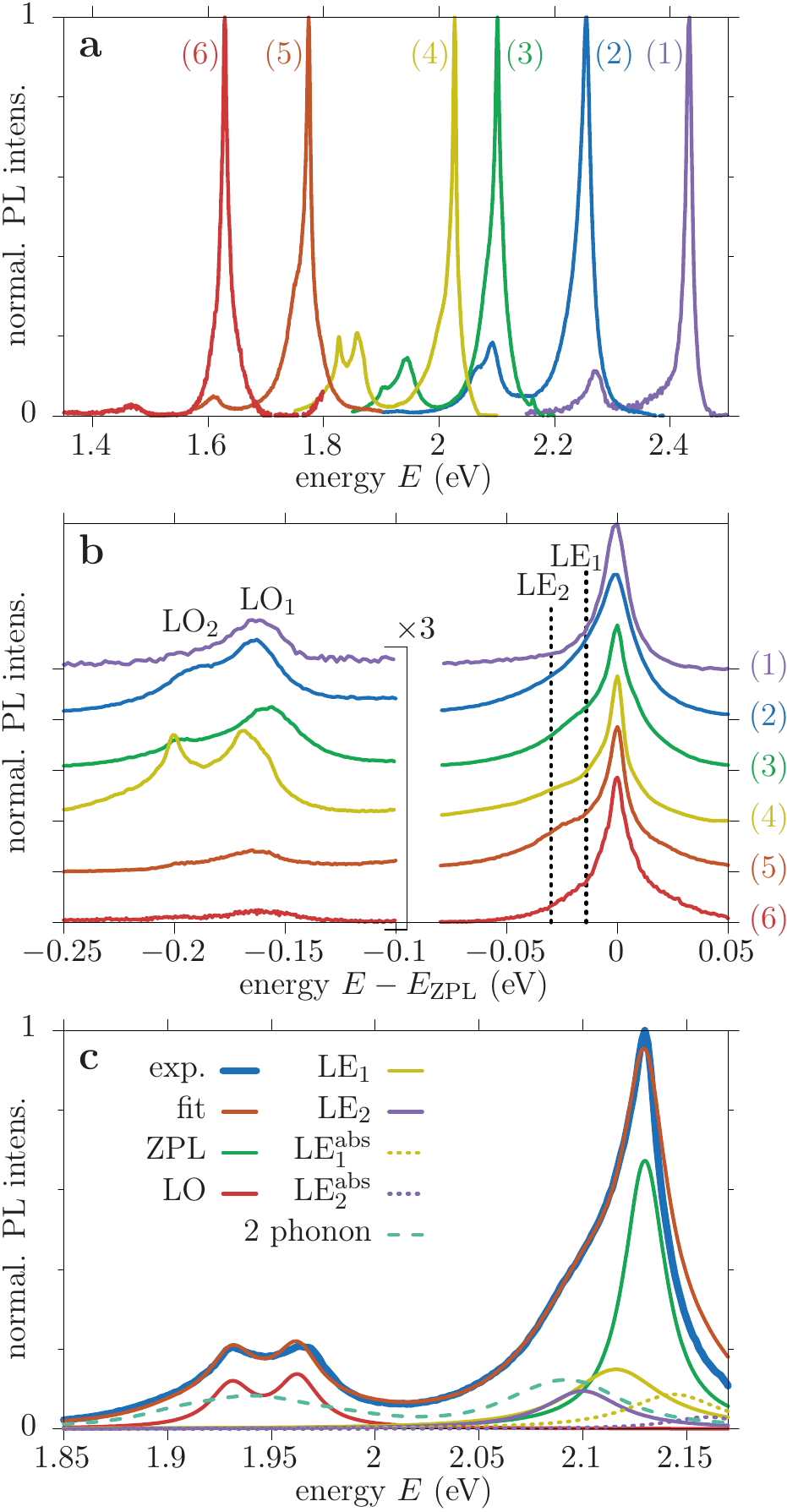}
\caption{{\bf Photoluminescence spectra.} (a) Exemplary room temperature photoluminescence (PL) spectra with different ZPL energies (1)--(6). (b) Same as (a) but plotted as a function of the detuning with respect to the ZPL energy to highlight the different shapes of the phonon side bands. (c) Phenomenological fit (orange) of a measured PL spectrum (blue). The constituents of the fit are the ZPL (green), single phonon LO side bands (red), the low energy (LE) side bands LE$_1$ (yellow) and LE$_2$ (violet) with the respective absorption peaks (dotted) and the two phonon side bands (dashed turquoise).}
\label{fig:1}
\end{figure}

Localized single-photon emitters are investigated in hBN nanopowder deposited on a Si/SiO$_2$ substrate (see Supplementary Material for details about the structure). Figure~\ref{fig:1} presents typical measured room-temperature PL spectra of six different localized light emitters in hBN. The transition energy of the emitters $E_{\rm ZPL}$, which is called the zero phonon line (ZPL) (labeled with numbers (1) to (6) in Fig.~\ref{fig:1}(a)) varies over a large range between 1.6~eV and 2.5~eV~\cite{tran2016rob}. The energy of the excited states of the emitters has already been extensively studied, both theoretically via DFT calculations~\cite{grinyaev2002dee,tawfik2017fir,abdi2018col,noh2018sta} and experimentally~\cite{tran2016rob,shotan2016pho,exarhos2017opt,li2017non,jungwirth2017opt,sontheimer2017pho,tran2017res,noh2018sta,xu2018sin,proscia2018nea}. Although different types of atomic defects such as N- or B-vacancies or substitutions with carbon or oxygen might exist, the distribution of transition energies is remarkably homogeneous. Therefore, it is unlikely that each emission energy in hBN stems from a different type of defect. One promising proposal for the measured wide range of transition energies is that at least part of the energy spread is due to a Stark shift~\cite{noh2018sta} resulting from an electric field which arises from a trapped charge near the color center~\cite{shotan2016pho}. Another suggested origin is a local strain distribution in the crystal~\cite{bourrellier2014nan,grosso2017tun,chejanovsky2017qua,proscia2018nea}. Our results suggest that both effects might contribute to the broad energy spread of the ZPL energies.

The narrow ZPLs of the emitters are accompanied by characteristic phonon side bands at lower energies. To compare the phonon side bands of the different emitters more easily, the spectra are plotted against the detuning from the ZPL energy in Fig.~\ref{fig:1}(b). Two side bands well separated from the ZPL between 0.2~eV and 0.15~eV below $E_{\rm ZPL}$ are visible, which can be attributed to two longitudinal optical (LO) phonons~\cite{martinez2016eff}. Also the ZPL itself is always more or less asymmetric with a steeper high energy side, which hints towards the contribution of low-energy (LE) phonon modes with energies below 50~meV. As will be shown later, we also assume two types of LE modes to be important. Comparing the phonon side bands of the different emitters in Fig.~\ref{fig:1}(b) we find significant variations. Not only the relative height of the LO phonon side bands compared to the ZPL intensity, but also the detailed structure, i.e., the relative heights of the LO$_1$ and LO$_2$ peak with respect to each other, vary from emitter to emitter. Having a closer look at the LE phonon modes near the ZPL, also this spectral phonon feature has different shapes and strengths for different emitters as highlighted in Fig.~\ref{fig:1}(b). In the following sections we will study these phonon side bands and develop a detailed understanding of their origin. In this context we will discuss which arguments support the identification of the LE phonons as longitudinal acoustic (LA) phonons and local mode oscillations of the defect with respect to the entire crystal.

\subsection{Phenomenological analysis}\label{sec:PL}
The properties of the LE and LO phonon side bands change significantly among the emitters. Therefore, we first try gaining a broad, quantitative overview of the different features. For this purpose we use a phenomenological model to fit the room temperature PL spectra of 165 different emitters. Figure~\ref{fig:1}(c) shows an exemplary spectrum (blue) and its phenomenological fit (orange). It is obvious that two LO phonon modes with energies at $E_{\rm LO,1}\approx 165$~meV and $E_{\rm LO,2}\approx 195$~meV have to be taken into account (red line). These energies are also supported by Raman measurements and {\it ab initio} calculations~\cite{serrano2007vib}. While trying different models for the LE phonons we found that at least two LE modes are required to achieve reasonable fits for all measured spectra. The best agreement was found for $E_{{\rm LE}_1} = 14$~meV and $E_{{\rm LE}_2}= 30$~meV. We will later show that these energies can be well identified as representative for LA phonons in the case of LE$_1$ and for a local mode in the case of LE$_2$. But for now we will stick to the label LE phonons for the low energy side bands. The applied fit function for the PL intensity reads
$$
I(E) = I_{\rm ZPL}(E) + I_{\rm LO}(E) + I_{\rm LE}(E)\ ,
$$
where we include a ZPL, single-phonon processes for the LO phonons and single- and two-phonon processes for the LE phonons. We also take single-phonon absorption processes for the LE modes into account. For all peaks we assume a Lorentzian shape. From each fit we retrieve the integrated peak weights $A_{\rm j}$ (${\rm j}= {\rm ZPL}$, LO$_1$, LO$_2$, LE$_1$, LE$_2$) and respective peak widths $\gamma_{\rm k}$ (${\rm k}= {\rm ZPL}$, LE$_1$, LE$_2$), where we assume that the LO side bands have the same broadening as the ZPL, which is a good approximation for dispersionless LO modes. Details about the fit function can be found in the Supplementary Material. In Fig.~\ref{fig:1}(c) we additionally show the two LE side bands (yellow for LE$_1$ and violet for LE$_2$) in phonon emission (solid) and absorption (dotted). The dashed turquoise line depicts the contribution from two phonon processes. Note that the absorption and two-phonon side bands are strictly related to the corresponding single-phonon emission side bands and therefore do not introduce additional fit parameters. The fit slightly deviates from the measurement for energies larger than the ZPL energy. The reason is, that a longpass filter is used close to the transition energy to block the exciting laser. This often reduces the measured high energy side of the ZPL. Therefore, we only used energies smaller than $E_{\rm ZPL}$ for the fits.

\begin{figure}[h]
\centering
\includegraphics[width=\columnwidth]{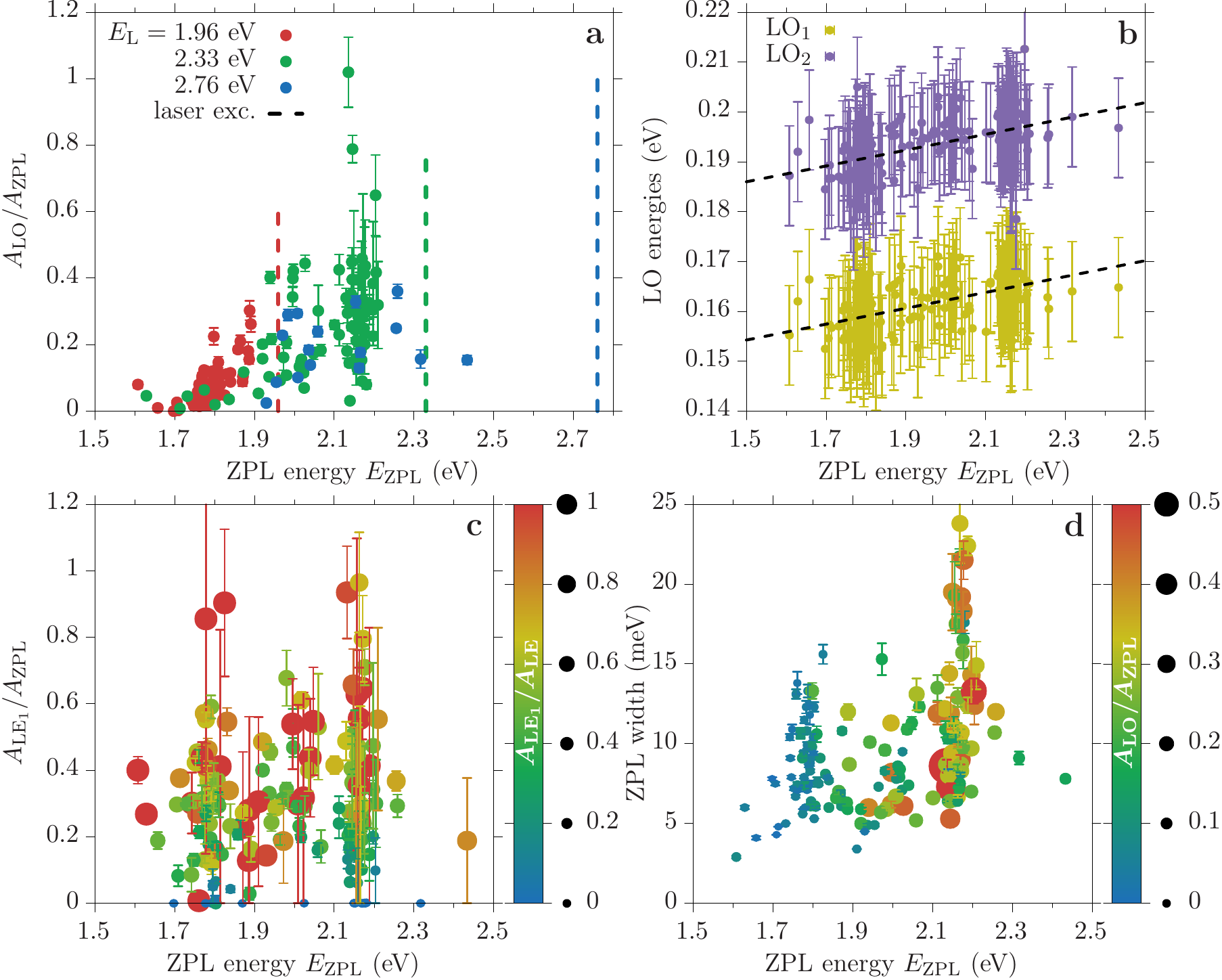}
\caption{{\bf Correlations of different phonon features.} (a) LO phonon side band ratio $A_{\rm LO}/A_{\rm ZPL}$ with the different dot colors indicating the excitation energy marked by the dashed lines. (b) Energy separation of the LO phonon side bands from the ZPL, LO$_1$ in yellow and LO$_2$ in violet. A constant error of 0.01~eV is assumed because these values are fixed by hand for every spectrum. (c) LE$_1$ phonon side band fraction $A_{\rm LE_1}/A_{\rm ZPL}$. The dot colors and sizes represent the contribution of LE$_1$ to the entire LE side band. (d) Width of the ZPL with the dot colors and sizes representing the LO side band fraction $A_{\rm LO}/A_{\rm ZPL}$ from (a).}
\label{fig:2}
\end{figure}

For the large data set of 165 emitters we identify correlations between the different spectral features. We start by comparing the LO phonon side bands to the ZPL. For this purpose, Fig.~\ref{fig:2}(a) shows the LO contribution, i.e., the ratio $A_{\rm LO}/A_{\rm ZPL}$ ($A_{\rm LO}=A_{{\rm LO}_1}+A_{{\rm LO}_2}$) between the LO side band and the ZPL weight, as a function of the ZPL energy. The colors of the dots represent the energy of the driving laser. The photon energies of the three lasers are marked by the dashed lines. We find transition energies almost equally distributed between 1.6~eV and 2.2~eV~\cite{jungwirth2016temp,shotan2016pho,li2017non,grosso2017tun,proscia2018nea}. Even a few emitters appear above 2.4~eV. Larger transition energies could not be measured, because of the longpass filter blocking the exciting laser. Having a look at the values of the LO contribution, i.e., the vertical distribution of the dots, we find that they have a clear tendency following a positive slope. While for transition energies below 1.9~eV most of the points lie between 0 and 0.2, they appear between 0.1 and 0.45 for energies around 2.1~eV. This shows that the emitters with smaller transition energies couple on average less efficiently to the optical phonons. This is in good agreement with the assumption that the shift of the ZPL energy is associated with the Stark effect. A static electric field acting on the emitter's dipole should increase the distance between the different charge centers. In consequence this should also influence the coupling to the polar LO phonons. The additional spread of LO side band strengths for a fixed energy, especially around 2.1~eV, could be due to the effect of local strain. Distortions of the lattice constants in the vicinity of the defect center should not only influence the distance between the charge centers, but also change the size of the emitter's wave function and therefore the effective distance between the charge centers. Later on we will discuss the influence of these properties on the phonon coupling in more detail.

To shed more light on the properties of the LO phonons, Fig.~\ref{fig:2}(b) shows the energies of the LO phonon side bands as a function of the ZPL energy. We find two clusters of data points, which we associate to the two LO phonon modes near the $\Gamma$ point. The energies of LO$_1$ (yellow) appear around 165~meV, those of LO$_2$ (violet) around 195~meV, which agrees well with calculated band structures and Raman measurements~\cite{serrano2007vib}. As indicated by the dotted lines, we find that the phonon energies have a slight trend to higher values for larger ZPL energies. As we identify these energies with the phonon energies of the hBN bulk crystal, this correlation suggests that the origin of the energy shift of the ZPL should also influence the LO phonon energies. This could be the case for a local strain field, which naturally changes the phonon dispersion. But also the dielectric environment might alter the LO phonon energies, as it affects in particular the LO-TO splitting~\cite{sohier2017bre}.

Figure~\ref{fig:2}(c) focusses on the LE phonon side bands by plotting the ratio between the first side band weight $A_{\rm LE_1}$ and the ZPL weight $A_{\rm ZPL}$ as a function of the ZPL energy. The dots are scattered over a wide range from 0 to almost 1 for any transition energy. However, the majority of the points is equally distributed between 0.1 and 0.7. This finding does not provide any strong correlation between the coupling to the LE$_1$ (LA) phonons and the transition energy. The color and size of the dots represents the LE$_1$ contribution to the entire LE side band, i.e., $A_{\rm LE_1}/A_{\rm LE}$, where $A_{\rm LE}=A_{\rm LE_1}+A_{\rm LE_2}$ (small dots are blue, large dots red). Also this quantity shows no correlation in the dot pattern. Small blue and large red dots are found everywhere. This suggests that the coupling to LE phonons is not strongly affected by the parameters that are expected to govern the transition energy, i.e., the distance of the charge centers. However, we find a wide spread of dots on the vertical axis, i.e., a non-negligible variation of coupling strengths for a given transition energy. This finding will be traced back to variations of the emitter's size as discussed in more detail later.

Finally, Fig.~\ref{fig:2}(d) presents the ZPL width as a function of ZPL energy. We find a slight trend to smaller line widths at small transition energies, which is a similar to what has been found in Fig.~\ref{fig:2}(a) for the strength of the LO side bands. This is in line with established models which associate the ZPL width with second-order phonon coupling mechanisms~\cite{machnikowski2006cha}. The mechanism suggested in Ref.~\cite{machnikowski2006cha} relies on the scattering of acoustic phonons with the polaron formed by optical phonons. Consequently, a weaker coupling to LO phonons should result in a smaller lattice displacement of the polaron and therefore in a smaller scattering rate for the acoustic phonons. This could lead to the narrowing of the ZPL we found here. Another reason for this trend could be a stronger spectral wandering induced by trapped charges which are also considered as the origin of the Stark shift of the ZPL energies~\cite{shotan2016pho}.

\subsection{Theoretical model}\label{sec:theory}
The phenomenological model used to fit the 165 spectra in Sec.~\ref{sec:PL} already includes some assumptions about the expected phonon modes, but it does not allow to retrieve further information about the emitters themselves. To support our assumptions about the mode energies we use a microscopically motivated model to reproduce the measured PL spectra. We also use this model to find possible origins for the correlations and non-correlations of the different phonon features and the ZPL energy found in Fig.~\ref{fig:2}. As already discussed in detail, our model should include the coupling to LO phonons to reproduce the two side bands around 165~meV and 195~meV below the ZPL. Also the coupling to acoustic phonons, which typically leads to a low-energy broadening of the ZPL, should be present in these systems. From defect centers in diamond, such as the extensively investigated NV$^{-}$ center, it is well known that the prominent phonon side band is dominated by the coupling to local mode oscillations~\cite{huxter2013vib,kehayias2013inf}. These modes describe displacements of the defect atoms with respect to the entire host crystal. Due to the similarity of the atomic defect structures, we expect that also in hBN local modes might have a significant influence on the PL spectrum.

We model the PL intensity spectrum of an emitter by calculating the optical susceptibility for a two-level system coupled to phonons, which  in the time domain is given by~\cite{duke1965pho,mahan2013man,schmitt1987the,krummheuer2002the}

\begin{equation}
\chi(t) = i\theta(t) \exp\left[i\omega_0 t + \Phi(t)-\frac{t}{T_2}\right]\ ,
\label{eq:chi}
\end{equation}
where $\theta(t)$ denotes the Heaviside step function, $\hbar\omega_0=E_{\rm ZPL}$ is the polaron-shifted transition energy of the emitter, i.e., the energy of the zero phonon line,  $T_2$ the dephasing time and $\Phi$ describes the phonon-induced dephasing with
\begin{equation}
\Phi(t) = \sum_{\rm j} \int\limits_{0}^{\infty} \frac{\mathcal{J}_{\rm j}(\omega_{\rm ph})}{\pi\omega_{\rm ph}^2} \Bigg\{\coth\left(\frac{\hbar\omega_{\rm ph}}{2k_{\rm B}T}\right)\Big[\cos(\omega_{\rm ph} t)-1\Big]-i\sin(\omega_{\rm ph} t)\Bigg\}  {\rm d}\omega_{\rm ph} \ .
\label{eq:Phi}
\end{equation}
This phonon influence is determined by the temperature $T$ and the phonon spectral density $\mathcal{J}_{\rm j}$ at the phonon energy $\hbar\omega_{\rm ph}$ for the phonon mode j. Note that such a model was commonly used to simulate the phonon coupling of F-centers~\cite{duke1965pho}. In this section $T=300$~K is considered to allow comparison with the experimental results in Sec.~\ref{sec:PL}. In the case of coupling to bulk phonons the spectral density is given by~\cite{reiter2014the}
\begin{equation}
\mathcal{J}_{\rm j}(\omega_{\rm ph}) = \sum_{\bf q} \left|g_{\rm j}({\bf q})\right|^2 \delta\big[\omega_{\rm ph}-\omega_{\rm j}({\bf q})\big]\ .
\label{eq:J}
\end{equation}
Here, $g_{\rm j}({\bf q})$ denotes the coupling matrix element for the coupling to a phonon with wave vector ${\bf q}$ in the phonon branch j and $\omega_{\rm j}({\bf q})$ is the corresponding phonon dispersion relation. With Eq.~\eref{eq:chi} the absorption spectrum $\alpha(\omega)$ of the emitter is simply given by the imaginary part of the Fourier transform
\begin{equation}
\alpha(\omega) =  {\rm Im}\left [\,\int\limits_{-\infty}^{\infty}\chi(t)\,{\rm e}^{-i\omega t}\,{\rm d}t\right]
\end{equation}
From this expression the emission spectrum $I(\omega)$ is then retrieved by inverting the absorption $\alpha$ with respect to the ZPL at $\omega_0$, i.e., 
\begin{equation}
I(\omega) = \alpha(-\omega + 2\omega_0)\ .
\end{equation}
We account for the coupling of the defect to LO$_1$ and LO$_2$ phonons via the Fr\"ohlich-interaction
\numparts
\begin{equation}
g_{\rm LO_j}({\bf q}) = \sqrt{\frac{e^2\omega_{\rm LO_j}({\bf q})}{2\varepsilon_0 V \hbar}\left[\frac{1}{\varepsilon_\infty(\bf q)}-\frac{1}{\varepsilon_{\rm s}(\bf q)}\right]} \,\frac{1}{q}\, \Big[ F_{\rm e}({\bf q}) -F_{\rm h}({\bf q})\Big]
\label{eq:LO}
\end{equation}
and for the coupling to LA phonons via the deformation potential coupling
\begin{equation}
g_{\rm LA}({\bf q}) = \frac{q}{\sqrt{2\varrho \hbar\omega_{\rm LA}({\bf q})V}} \, \Big[ D_{\rm e}F_{\rm e}({\bf q}) - D_{\rm h}F_{\rm h}({\bf q})\Big]\ ,
\label{eq:LA}
\end{equation}
\endnumparts
where $V$ is a normalization volume, $\varrho$ is the mass density, $\varepsilon_{\rm s}$ and $\varepsilon_\infty$ are the static and high frequency dielectric functions, respectively, and $D_{\rm e}$ and $D_{\rm h}$ are the deformation potentials for electrons and holes, respectively. The form factors $F_{\rm e,h}({\bf q})$ are determined by the wave function of the excited state of the emitter as we explain in the following.

Another common coupling mechanism between charges and LA phonons is via the piezoelectric effect. In Ref.~\cite{michel2011pho} it was shown that the piezoelectric constants vanish with an increasing number of hBN layers. As we are investigating nanopowder samples, we are dealing with small bulk crystals with large layer numbers. The thickness of the investigated hBN crystals is continuously distributed between 40 nm and 100 nm as exemplarily shown in the Supplementary Material. In Ref.~\cite{michel2011pho} it was demonstrated that due to inversion symmetry the piezoelectric constants vanish for even layer numbers. If this coupling mechanism to LA phonons played an important role, it should only appear for half of the studied defects, i.e., those located in samples with odd layer numbers. This effect should lead to a separation of data points for crystals of even and odd layer number in Fig.~\ref{fig:2}(a) but one does not find such a feature. Therefore, we conclude that the piezoelectric coupling to LA phonons does not play an important role in our samples and assume that this coupling mechanism can be neglected.

\begin{figure}[h]
\centering
\includegraphics[width=0.6\columnwidth]{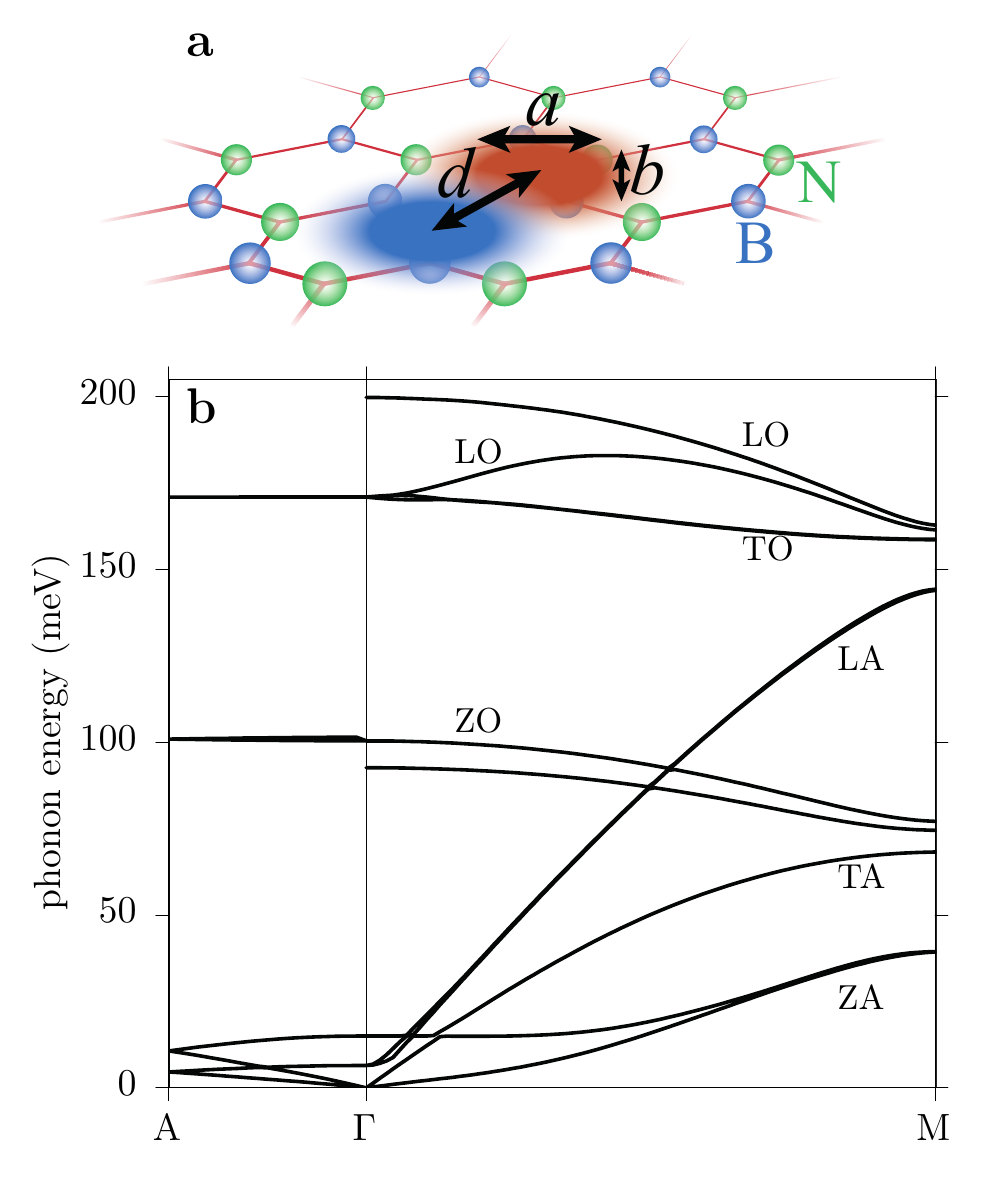}
\caption{{\bf Model system.} (a) Schematic drawing of the wave function of the defect in a single hBN layer. The charge centers have a Gaussian shape with an in-plane localization length $a$, an out-of plane localization length $b$ and an in-plane distance $d$. (b) Phonon dispersion of hBN adapted from Ref.~\cite{serrano2007vib}. The A$\Gamma$ direction is the out-of-plane direction and the $\Gamma$M direction an example for an in-plane direction.}
\label{fig:3}
\end{figure}

One central aspect of hBN is its pronounced anisotropy arising from its layered structure. In our model we take this into account by distinguishing in-plane and out-of-plane directions. We assume that the dipole of the emitter lies in the plane of one hBN layer. Therefore, we consider a wave function of the defect's excited state consisting of two differently charged centers with an in-plane distance $d$ as schematically shown in Fig.~\ref{fig:3}(a). The negative charge will be referred to as electron~(e) and the positive one as hole~(h). Each of the charge centers has a Gaussian wave function with an in-plane localization length $a$ and an out-of-plane localization length $b$. This form of the wave function can be seen as an approximation for a wave function typically calculated from {\it ab initio} theory. The advantage of our approach is the flexibility to easily study the influence of changes of geometrical quantities. With this form of the wave function the form factors in Eqs.~\eref{eq:LO} and~\eref{eq:LA} read
\numparts
\begin{eqnarray}
F_{\rm e}({\bf q}) &=& \exp\left[-\left(\frac{q_r a}{2}\right)^2-\left(\frac{q_z b}{2}\right)^2 -  \frac{i}{2} {\bf d}\cdot {\bf q}\right]\qquad {\rm and}\label{eq:F_e}\\
F_{\rm h}({\bf q}) &=& \exp\left[-\left(\frac{q_r a}{2}\right)^2-\left(\frac{q_z b}{2}\right)^2 + \frac{i}{2} {\bf d}\cdot {\bf q}\right]\ , \label{eq:F_h}
\end{eqnarray}
\endnumparts
where the phonon wave vector ${\bf q}$ is split into an in-plane $q_r$ and an out-of-plane $q_z$ component and ${\bf d}$ is the vector connecting the two charge centers. After integrating $|g_{\rm j}({\bf q})|^2$ in Eq.~\eref{eq:J} over the in-plane angle of the phonon wave vector, the influence of the wave function geometry on the coupling strengths is given by
\numparts
\begin{equation}
\big|g_{\rm LO_j}(q_r,q_z)\big|^2 \sim \exp\left\{-\frac12\Big[\left(q_r a\right)^2 + \left(q_z b\right)^2\Big]\right\} \Big[1-J_0(dq_r)\Big]
\label{eq:gLO}
\end{equation}
and
\begin{equation}
\big|g_{\rm LA}(q_r,q_z)\big|^2 \sim \exp\left\{-\frac12\Big[\left(q_r a\right)^2 + \left(q_z b\right)^2\Big]\right\} \Big[D_{\rm e}^2+D_{\rm h}^2-2D_{\rm e}D_{\rm h}J_0(dq_r)\Big]\ ,
\label{eq:gLA}
\end{equation}
\endnumparts
where $J_0$ is the Bessel function of first kind and of zeroth order. Due to these form factors, coupling to phonons is essentially restricted to the wave vector region $q_r \lesssim a^{-1}$, $q_z \lesssim b^{-1}$.

For the Fr\"ohlich coupling, different dielectric constants for the in-plane and the out-of-plane direction are considered. The energies of the phonons, which enter the model, are extracted from the simulated dispersion relation shown in Fig.~\ref{fig:3}(b) adapted from Ref.~\cite{serrano2007vib}; we assume two LO modes (LO$_1$ and LO$_2$) with constant dispersions. For the LA phonons we consider linear dispersions, where we again distinguish between the in-plane and the out-of-plane direction. More details about the model parameters can be found in the Supplementary Material. As will be shown below, the 14~meV LE$_1$ mode introduced in Sec.~\ref{sec:PL} can be seen as a reasonable approximation for the side band stemming from the LA phonons. In Ref.~\cite{tawfik2017fir} it was calculated that the C$_{\rm B}$N$_{\rm V}$ center has the strongest coupling to an in-plane breathing mode of the defect center, which has an energy of 30~meV. Also other studies suggest this atomic structure as promising candidate for the hBN color center~\cite{sajid2018def,lopez2018tow}. This motivates us to interpret the LE$_2$ mode in the phenomenological model in Sec.~\ref{sec:PL} as representative of such a breathing mode. The coupling to this local mode is modeled by a Lorentzian-like distribution of the phonon spectral density~\cite{garg1985eff,thorwart2000con,norambuena2016mic}
 \begin{equation}
\mathcal{J}_{\rm LOC}(\omega_{\rm ph}) = \frac{1}{\pi}\frac{16 g^2 \omega_{\rm LOC}\Delta \omega_{\rm ph}}{\left(\omega_{\rm ph}^2-\omega_{\rm LOC}^2\right)^2+4 \Delta^2\omega_{\rm ph}^2}\ .
\label{eq:J_loc}
 \end{equation}
The parameters entering this coupling are the local mode energy $\hbar\omega_{\rm LOC}=E_{\rm LOC}$, the coupling strength $g$ and the width $\Delta$. For the considered parameters the spectral shape of the side band due to the local mode agrees well with the ordinary Lorentzian shape considered for the phenomenological fits in Sec.~\ref{sec:PL}. We show in the Supplementary Material that our model for the local mode coupling reproduces PL spectra of different defect centers in diamond very well, showing the accuracy of the model. To reproduce the finite width of the ZPL, we assume a constant dephasing rate $1/T_2$, providing us with a dephasing time $T_2$. We note that this dephasing time includes pure dephasing, homogeneous broadening mechanisms as well as inhomogeneous processes, e.g., spectral wandering~\cite{sontheimer2017pho}.

While the phonon dispersion relations and dielectric constants are well known for hBN, the deformation potentials are only hardly known. Simulations indicate that the relevant values are in the range of some hundred meV to a few eV~\cite{wiktor2016abs}. We therefore will try finding reasonable values for the deformation potentials $D_{\rm e}$ and $D_{\rm h}$ for the LA phonon coupling of the considered defect centers in hBN. The other unknown parameters in our model are the in-plane and out-of-plane localization lengths $a$ and $b$, respectively, and the charge separation $d$ in the defect state. To be able to extract information on these parameters from the measured spectra, we will analyze their role for the coupling to LO and LA phonon modes, respectively.

\begin{figure}[h]
\centering
\includegraphics[width=\columnwidth]{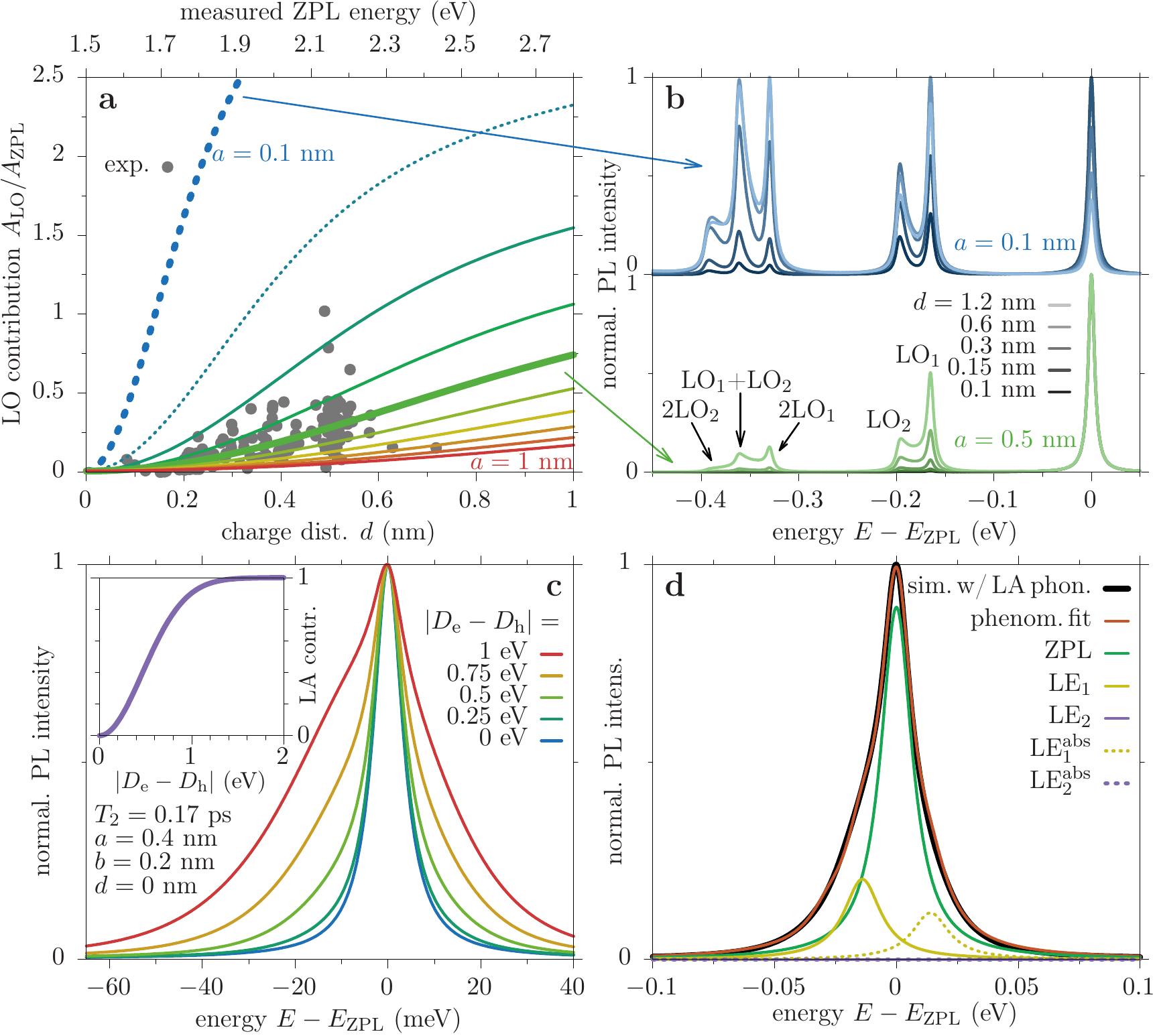}
\caption{{\bf } (a) Contribution of the first LO phonon side band with respect to the ZPL $A_{\rm LO}/A_{\rm ZPL}$ as a function of the charge distance $d$ for different localization lengths $a$. The gray dots are the measurements from Fig.~\ref{fig:2}(a). (b) Exemplary simulated normalized PL spectra for different $d$ from large (bright) to small values (dark), for $a=0.1$~nm (top) and $a=0.5$~nm (bottom). (c) Exemplary simulated normalized PL spectra for different values of $|D_{\rm e}-D_{\rm h}|$ and vanishing charge separation. Inset: LA phonon contribution as a function of deformation potential difference $|D_{\rm e}-D_{\rm h}|$. (d) Phenomenological fit of a simulated PL spectrum including the coupling to LA phonons.}
\label{fig:4}
\end{figure}

For the coupling to LO phonons the only unknown quantities are the geometrical parameters in-plane localization length $a$, out-of-plane localization length $b$, and charge distance $d$. To investigate their influence, we first concentrate on the coupling to LO phonons. In Fig.~\ref{fig:4}(a) we plot the contribution of the LO single-phonon side band, i.e., the fraction between the weight of the first LO phonon side bands of LO$_1$ and LO$_2$ and the ZPL weight, as a function of the charge distance $d$ for different localization lengths $a$ between 0.1~nm (blue) and 1~nm (red). The distance $d$~is varied on the same scale. This should be a reasonable range for these quantities as the in-plane lattice constant of hBN is 0.4~nm and the extension of an atomistic defect state should be of similar size. Under the assumption that the Stark effect is responsible for the different transition energies of the emitters, we additionally plot the measured data points from \ref{fig:2}(a) as gray dots into Figs.~\ref{fig:4}(a). The relation between the energies of the experimental values of Fig.~\ref{fig:2}(a) (gray dots in Fig.~\ref{fig:4}(a)) and the charge distances $d$ in Fig.~\ref{fig:4}(a) is obtained by the fits of the measured spectra in Sec.~\ref{sec:compare}, which yield values between $d = 0.2$~nm and $d=0.5$~nm. By the given choice of the ZPL energy axis the majority of the measurement points appear in this range. The simulated LO contributions vanish for vanishing in-plane distance $d$ because in this limit there is no dipole moment associated with the defect. For a given in-plane localization length $a$ the LO contribution grows with increasing $d$. When increasing the localization length $a$ the curves rise slower, because the Fr\"ohlich coupling strength depends on the effective dipole strength of the wave function. To achieve the same strength for larger localization lengths $a$ the charge separation has to be accordingly larger. The increase of the LO contribution with increasing distance nicely resembles the distribution of the measured dot pattern, which supports the interpretation that the shift of the ZPL is at least partially related to a Stark shift. To further analyze the influence of the LO phonons, we show exemplary normalized PL spectra for different values of $a$ and $d$ in Fig.~\ref{fig:4}(b). The colors agree with the corresponding curves in (a). At the bottom for $a=0.5$~nm the two single-phonon LO side bands are well resolved around $-0.2$~eV, also the three two-phonon side bands between $-0.3$~eV and $-0.4$~eV can be seen. As expected the amplitude of the side bands increases for larger in-plane distance $d$. At the top of Fig.~\ref{fig:4}(b) the same spectra are shown for $a=0.1$~nm. Here, the LO side bands grow very rapidly when increasing $d$, they even become larger than the ZPL. For the largest considered $d$ the two-phonon side bands are even larger than the single-phonon side bands. Indeed, from Eqs.~\eref{eq:Phi} and \eref{eq:J} it can be shown that, as long as successive side bands do not spectrally overlap, the contributions of single-phonon ($A_{\rm LO}$) and two-phonon ($A_{\rm 2LO}$) side bands with respect to the ZPL are given by
\begin{equation}
\frac{A_{\rm LO}}{A_{\rm ZPL}} = \sum_{{\rm j},{\bf q}} \left| \frac{g_{\rm LO_j}({\bf q})}{\omega_{\rm LO_j}} \right|^2 \ , \qquad \frac{A_{\rm 2LO}}{A_{\rm ZPL}} =\frac{1}{2} \left( \frac{A_{\rm LO}}{A_{\rm ZPL}}\right)^2 \ .
\end{equation}
Thus, as soon as $A_{\rm LO}>2A_{\rm ZPL}$ the two-phonon sideband will exceed the single-phonon one. However, in the measured PL spectra of hBN color centers this regime where the side bands exceed the ZPL is not reached. Therefore, we can conclude that the localization length should lie between $a=0.3$~nm and 1~nm and the charge distance between $d=0$~nm and 1~nm. 

After finding reasonable values for the in-plane size of the wave function ($a$ and $d$), in Figs.~\ref{fig:4}(c) we focus on the values for the deformation potentials $D_{\rm e}$ and $D_{\rm h}$. When considering the special case of a vanishing charge distance, $d=0$, the deformation potential coupling strength in Eq.~\eref{eq:gLA} is proportional to $(D_{\rm e}-D_{\rm h})^2$. So we will use this limit to find reasonable values for the difference of the deformation potentials. To provide a quantitative measure, we calculate the LA contribution to the entire spectrum via
\begin{equation}
1-{\rm Im}\bigg[ \chi_{\rm LA}(t\to \infty; \omega_0=0; T_2\to\infty) \bigg]\ .
\end{equation}
This value is plotted as a function of $|D_{\rm e}-D_{\rm h}|$ in the inset in Fig.~\ref{fig:4}(c) for wave function dimensions of $a=0.4$~nm, $b=0.2$~nm and $d=0$~nm. For $D_{\rm e}=D_{\rm h}$ the LA contribution vanishes and no LA side band appears in the spectrum, while the side band dominates the entire spectrum for large values of $|D_{\rm e}-D_{\rm h}|$. In Fig.~\ref{fig:2}(c) we found that the LA contribution in the measured spectra also for small transition energies, i.e., small $d$, yields values between zero and one. Therefore, reasonable values for the difference of the deformation potentials should lead to an LA contribution of around 0.5. We find that this is fulfilled for $|D_{\rm e}-D_{\rm h}|\approx 0.6$~eV. However, the deformation potentials not only determine the LA contribution but are also responsible for the shape of the spectra. To demonstrate this influence, we show exemplary spectra for different values of $|D_{\rm e}-D_{\rm h}|$ still for a vanishing dipole moment $d=0$ in Fig.~\ref{fig:4}(c). We find that with increasing $|D_{\rm e}-D_{\rm h}|$ (blue to red) the asymmetry of the ZPL peak increases till the entire spectrum is dominated by the LA side band for the 1~eV case (red). Here the ZPL is only a rather small peak on the broad phonon background. From here on we fix $|D_{\rm e}-D_{\rm h}|=0.6$~eV.

As already mentioned, we assume that the coupling to the LE$_1$ mode phenomenologically describes the coupling to LA phonons. One might wonder why the coupling to a discrete mode with a Lorentzian lineshape approximates reasonably well the coupling to the continuum of LA phonons. To confirm that this is indeed a valid approximation, the black solid line in Fig.~\ref{fig:4}(d) shows a simulated room-temperature PL spectrum, where the coupling to the continuum of LA phonons is fully taken into account. This spectrum was then fitted by the phenomenological model of Sec.~\ref{sec:PL} yielding the orange line. The contributions of the two LE modes (yellow and violet) and the ZPL (green) in the phenomenological model are plotted separately as before. We find an excellent agreement between the simulated spectrum including only LA phonons and the phenomenological fit. Furthermore we find that only the LE$_1$ mode has a relevant contribution in the fit. This shows that indeed the LE$_1$ mode is a reasonable approximation for the contribution of the LA phonons to the spectrum.

\begin{figure}[h]
\centering
\includegraphics[width=\columnwidth]{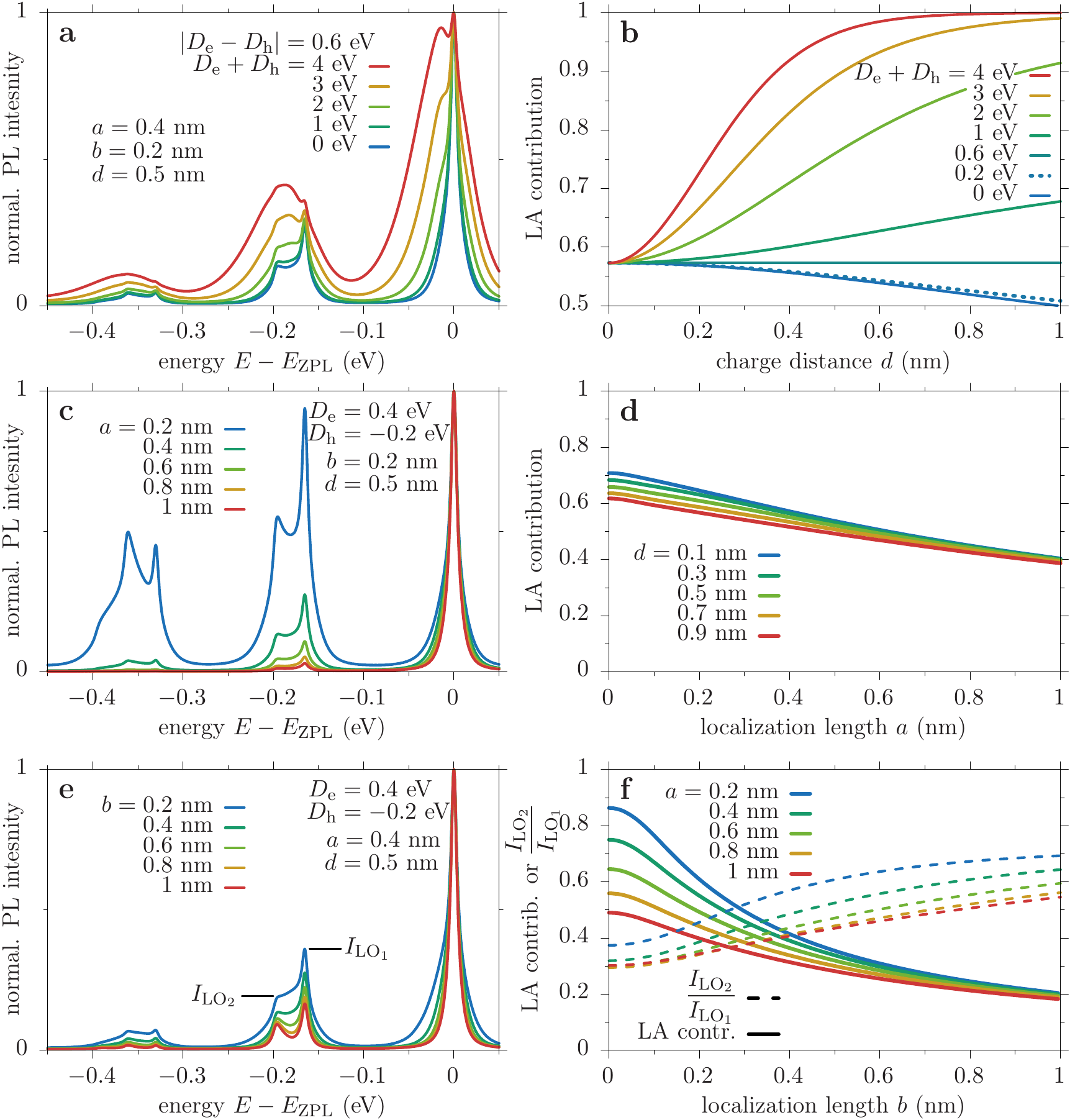}
\caption{{\bf } Influence of (a,b) deformation potentials $D_{\rm e}+D_{\rm h}$ and charge separation $d$, (c,d) in-plane localization length $a$ and (e,f) out-of-plane localization length $b$ on the spectra. (a,c,e) Exemplary normalized PL spectra. (b,d,e) LA contribution as functions of the wave function dimensions $d$, $a$ and $b$. (f) also shows the fraction of the LO single-phonon side band peak intensities.}
\label{fig:5}
\end{figure}

In the next step we investigate the influence of the sum of the deformation potentials. For a non-vanishing charge distance $d$ the Bessel function in Eq.~\eref{eq:gLA} does not vanish leading to the fact that the phonon coupling depends on $D_{\rm e}$ and $D_{\rm h}$ separately. In Fig.~\ref{fig:5}(a) we show simulated PL spectra for $a=0.4$~nm, $b=0.2$~nm and $d=0.5$~nm and different values of $D_{\rm e}+D_{\rm h}$ between 0~eV (blue) and 1~eV (red). We find that the weight of the LA side bands close to the ZPL strongly increases with the sum of the deformation potentials. It is also nicely visible that the LO phonon side bands are additionally broadened by the LA phonons, demonstrating that multi-phonon contributions are well reproduced by the model. In Fig.~\ref{fig:2}(c) we found that the LA contribution (LE$_1$ mode) to the spectrum exhibits no significant correlation with the transition energy, i.e., with $d$. This finding should be reproduced by the choice of $D_{\rm e}+D_{\rm h}$. Therefore, in Fig.~\ref{fig:5}(b) we plot the LA contribution as a function of the charge distance $d$ for different values of $D_{\rm e}+D_{\rm h}$ between 0~eV (blue) and 4~eV(red) ($a=0.4$~nm and $b=0.2$~nm). All curves start at the same value of slightly below 0.6 for $d=0$~nm. We find that for deformation potential sums of 1~eV or larger the LA contribution grows significantly with increasing $d$. One would therefore expect a clear correlation between these two quantities. This makes us conclude that $D_{\rm e}+D_{\rm h}$ should be smaller than 1~eV. In the special case of $D_{\rm e}+D_{\rm h}=0.6~{\rm eV}=|D_{\rm e}-D_{\rm h}|$ the LA contribution is a straight line, because in this case either $D_{\rm h}=0$~eV or $D_{\rm e}=0$~eV, which makes the charge distance irrelevant. However, we do not expect one of the  deformation potentials, either for electrons or for holes, to vanish in hBN. This would mean that either electrons or holes do not scatter with LA phonons via the deformation potential coupling, which makes us disregard this value. We found good results for $D_{\rm e}+D_{\rm h}=0.2$~eV ($D_{\rm e}=0.4$~eV, $D_{\rm h}=-0.2$~eV) as will be demonstrated below in Sec.~\ref{sec:compare} when directly comparing the simulations with measured PL spectra.

After having selected reasonable values for the deformation potentials, we now study the influence of the other two geometrical parameters, the in-plane localization length~$a$ and the out-of-plane localization length $b$ of the wave function. Figures~\ref{fig:5}(c,d) focus on the in-plane localization length $a$. In Fig.~\ref{fig:5}(c) exemplary spectra for $d=0.5$~nm, $b=0.2$~nm and different values of $a$ between 0.2~nm (blue) and 1~nm (red) are shown. While the LO phonon side bands increase significantly when reducing $a$, as discussed earlier, the LA phonon side band next to the ZPL does not change significantly. To provide a quantitative measure for the influence of $a$ on the LA coupling in Fig.~\ref{fig:5}(d) we plot the LA contribution as a function of $a$ for different values of $d$. All curves slightly decrease from values between 0.6 and 0.7 to 0.4 for $a=1$~nm. This spread can give rise to part of the spread of LA contributions found in Fig.~\ref{fig:2}(c).

Finally, the influence of the out-of-plane localization length $b$ is considered. In Fig.~\ref{fig:5}(e) we again show PL spectra for $a=0.4$~nm, $d=0.5$~nm and different values of $b$ between 0.2~nm (blue) and 1~nm (red). Two pronounced features are found in the spectra: (i) the LA phonon side band increases with decreasing $b$; (ii) the two LO single-phonon side bands LO$_1$ and LO$_2$ slightly increase with decreasing $b$, but not by the same rate. While LO$_1$ and LO$_2$ are almost equally strong for $b=1$~nm (red), LO$_1$ is significantly stronger for 0.2~nm (blue). These two features are quantified in Fig.~\ref{fig:5}(f). The solid lines show the LA contribution and the dashed lines the ratio between the peak intensities of the two LO single-phonon side bands $I_{\rm LO_2}/I_{\rm LO_1}$ as functions of $b$ for different values of $a$. The LA contribution rapidly drops from values up to 0.9 for $b=0$~nm to 0.2 for $b=1$~nm. This spread of LA phonon side band contributions agrees well with the measured values found in Fig.~\ref{fig:2}(c). The fraction of the LO side band peaks increases for growing $b$. This variation of the shape of the LO side band was also found in the measured spectra in Fig.~\ref{fig:1}. The reason for this variation of the LO side bands is that the larger LO phonon energy is not present for the out-of-plane phonon wave vectors, as can be seen in Fig.~\ref{fig:3}(b). So $b$ mainly influences the LO$_1$ phonon side band. A larger out-of-plane localization length $b$ in Eq.~\eref{eq:gLO} leads to a smaller range of out-of-plane phonon wave vectors that is excited by the emitter, which leads to a weaker effective coupling strength. Therefore the fraction $I_{\rm LO_2}/I_{\rm LO_1}$ reduces for growing $b$. The out-of-plane lattice constant of hBN is 0.66~nm, which makes the considered values for $b$ a reasonable range for this parameter.

This parameter study shows that slight variations of the geometrical parameters of the defect's wave function in the range of reasonable dimensions lead to strong changes in the PL spectra. In the next step, we apply this model to directly reproduce measured PL spectra by varying the geometrical parameters. We also take the local mode for the low-energy coupling in the range of 30~meV into account and vary its coupling parameters.

\subsection{Comparison between experiment and theory}\label{sec:compare}

\begin{figure}[h]
\centering
\includegraphics[width=0.8\columnwidth]{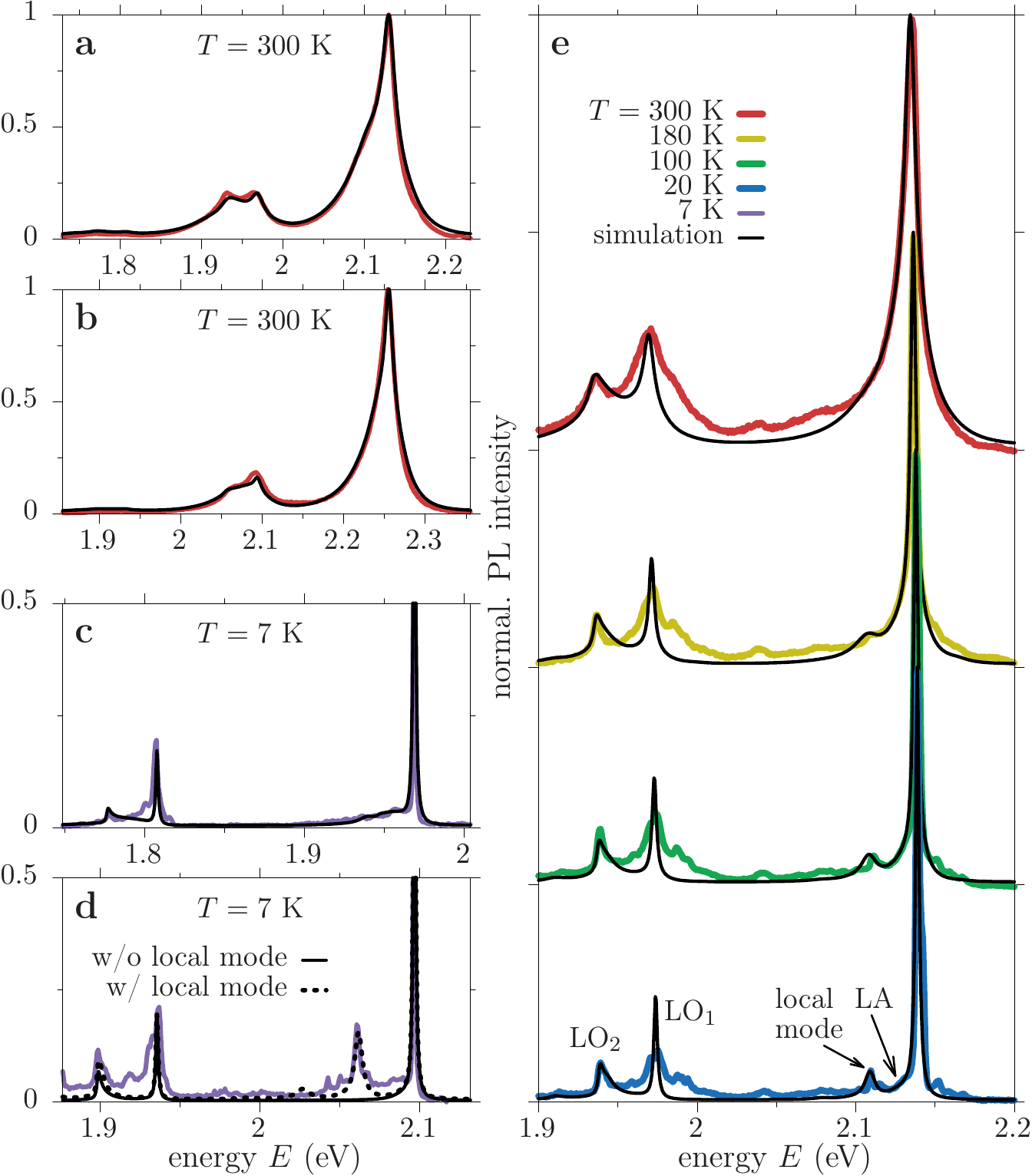}
\caption{{\bf Reproducing measured PL spectra.} Normalized measured (colored) and simulated (black) PL spectra. (a,b) room temperature 300~K, (c,d) cryogenic temperature 7~K. (e) Temperature series from 300~K to 20~K from top to bottom of the same emitter. The model parameters are given in Tab.~\ref{tab:1}.}
\label{fig:6}
\end{figure}

Figure~\ref{fig:6} shows PL measurements of different defect centers. While the spectra in Figs.~\ref{fig:6}(a,b) are taken at room temperature, Figs.~\ref{fig:6}(c,d) are measured at cryogenic temperatures of about 7~K. Figure~\ref{fig:6}(e) depicts a temperature series taken on the same emitter. The temperature decreases from top to bottom from 300~K to 20~K. In addition to the experiments (colored lines) we show corresponding simulations as black lines. All parameters for the simulations are listed in Tab.~\ref{tab:1}.

\begin{table}[h]
\caption{\label{tab:1}Model parameters for the simulations in Fig.~\ref{fig:6} including the temperature $T$, the localization lengths $a$ and $b$, the charge distance $d$, the LO phonon energies $E_{\rm LO_1}$ and $E_{\rm LO_2}$, and the local mode energy $E_{\rm LOC}$ and coupling strength $g/\omega_{\rm LOC}$. The local mode width $\hbar\Delta$ and the dephasing time $T_2$ depend on the temperature.}
\begin{tabular}{@{} c | c c c c c c c c c c}
\br
Fig.~\ref{fig:6}	& $T$ 	& $a$ 	& $b$ 	& $d$	& $E_{{\rm LO}_1}$ 	& $E_{{\rm LO}_2}$ 	&  $E_{\rm LOC}$ 	& $g/\omega_{\rm LOC}$ 	& $\hbar\Delta$		& $T_2$\\
			& (K) 	& (nm)	& (nm) 	& (nm) 	& (meV) 			& (meV) 			& (meV) 			&		& (meV) 		& (ps)\\
\mr
(a) & 300	& 0.40	& 0.40	& 0.44	& 165 	& 200.0 	& 30	& 0.38	& 15 		& 0.08 \\
(b) & 300	& 0.50	& 0.10	& 0.40	& 163 	& 200.0 	& 30	& 0.30	& 18 		& 0.12 \\
(c) &     7	& 0.40	& 0.15	& 0.25	& 161 	& 192.0 	& 30	& 0.15	& 5.0 	& (0.8) \\
(d) &     7	& 0.20	& 0.40	& 0.22	& 161 	& 198.0 	& (35)& (0.35)	& (2.0)	& (0.8) \\
(e) & 300	& 0.25	& 0.30	& 0.30	& 165 	& 200.5 	& 30	& 0.21	& 15 		& 0.17 \\
     & 150	&         	&        	&        	&   		& 		& 	&     		& 7.0 	& 0.33 \\
     &   50	&         	&         	&        	&   		&  		& 	&     		& 4.0 	& (0.4) \\
     &   20	&         	&         	& 	  	&   		&  		& 	&     		& 2.5 	& (0.5) \\
\br
\end{tabular}
\end{table}

Comparing the different spectra, we find a convincing agreement between the measured and the calculated spectra. The theoretical approach used in Ref.~\cite{exarhos2017opt}, which considers the phonon spectral function as a free parameter, also led to good agreements with measured spectra. However, the authors did not draw conclusions on the emitters' wave functions and details of the electron-phonon coupling matrix elements. From the parameters for the dimensions of the wave functions in our model listed in Tab.~\ref{tab:1} we find that all sizes ($a$, $b$, and $d$) are in the range between 0.1~nm and 0.5~nm. This is a reasonable parameter range, because it shows that the assumed emitter wave function spreads over a few unit cells at most and is mainly restricted to the layer in which the defect is located. While the LO phonon energies scatter slightly around the already discussed values of $E_{{\rm LO}_1}\approx 165$~meV and $E_{{\rm LO}_2}\approx 195$~meV, the energy of the local mode is rather fixed. This might have two reasons. On the one hand, the side band in the PL spectrum partly overlaps with the LA contribution at room temperature, which makes a precise determination of the peak position less easy. On the other hand, strain distributions could have less influence on the energetic position of the mode, but rather change the coupling strength $g$ and the width $\Delta$. We want to remark that, the low temperature spectrum in Fig.~\ref{fig:6}(d) shows a pronounced peak at $E\approx 2.05$~eV, i.e., 35~meV below the ZPL. The dashed line includes this peak as a local mode in the simulation. However, this spectral feature could also stem from another nearby emitter. Usually, PL spectra at cryogenic temperatures exhibit a large number of narrow lines in these samples, which makes a clear identification of the lines difficult.

To reproduce the entire temperature series in Fig.~\ref{fig:6}(e) we fix the dimensions of the wave function and the local mode coupling strength for $T=300$~K as given in Tab.~\ref{tab:1}. For each temperature we have to additionally determine the local mode width $\Delta$ and the dephasing time $T_2$. We find that both the ZPL and the local mode side band, which remains as narrow peak slightly above $E=2.1$~eV for $T=20$~K, become narrower with lower temperatures. This can already be seen when comparing the room-temperature spectra in Figs.~\ref{fig:6}(a,b) to the low-$T$ ones in Figs.~\ref{fig:6}(c,d). In the Supplementary Material we discuss another temperature series of PL spectra taken on a different emitter, where we additionally measured the PL lifetime. In agreement with Ref.~\cite{jungwirth2016temp} we find that the life time does not depend on the temperature and that it is in the range of a few nanoseconds~\cite{tran2016quaI,tran2016quaII,tran2016rob,martinez2016eff,sontheimer2017pho,proscia2018nea}. This indicates that the ZPL line width at room temperature is dominated by the pure dephasing mechanism and spectral wandering, which we combine in the fit parameter of the dephasing time $T_2$. For decreasing temperatures, the ZPL width shrinks significantly as also found in Refs.~\cite{vuong2016pho,jungwirth2016temp}. However, the values for low temperatures between 5 K and 50 K are underestimated because the measured line width is on the order of the limited resolution of the spectrometer (100 \textmu eV). Hence, we are not able to draw reliable conclusions about the low-temperature $T_2$ times from our data. The evolution of the two low-energy phonon side bands with decreasing temperature nicely supports our assumptions of their nature. The local mode results in an isolated peak at the same energy. This is not expected for phonons with continuous spectrum, i.e., for LA modes. The LA phonons lead to a low-energy side band directly attached to the ZPL, as it is well known from semiconductor QDs~\cite{jakubczyk2016imp}. In the spectra in Fig.~\ref{fig:6}(e) we find that the spectral shape of the LO$_1$ phonon side band is additionally broadened to the high energy side compared to the simulation. When moving from high to low $T$, all spectral features become narrower, which makes it obvious that this broadening evolves into another maximum just below $E=2$~eV. The fits for the low-temperature spectra in Figs.~\ref{fig:6}(c,d) show that our model reproduces the LO phonon side bands, despite our rough approximations for the phonon dispersion when assuming constant LO phonon energies. However, it was shown that strain strongly influences the emitters' spectra~\cite{bourrellier2014nan,grosso2017tun,chejanovsky2017qua} and that it is highly likely that the emitters appear near the surface of the sample~\cite{shotan2016pho, tran2016rob}. Local strain or the proximity to the surface will alter the phonon band structure and could thereby lead to variations of the spectral shape of the LO side bands. This could be a reason for the atypical broadening of the LO$_1$ side band for this emitter.

\subsection{Photoluminescence excitation spectra}

\begin{figure}[h]
\centering
\includegraphics[width=\columnwidth]{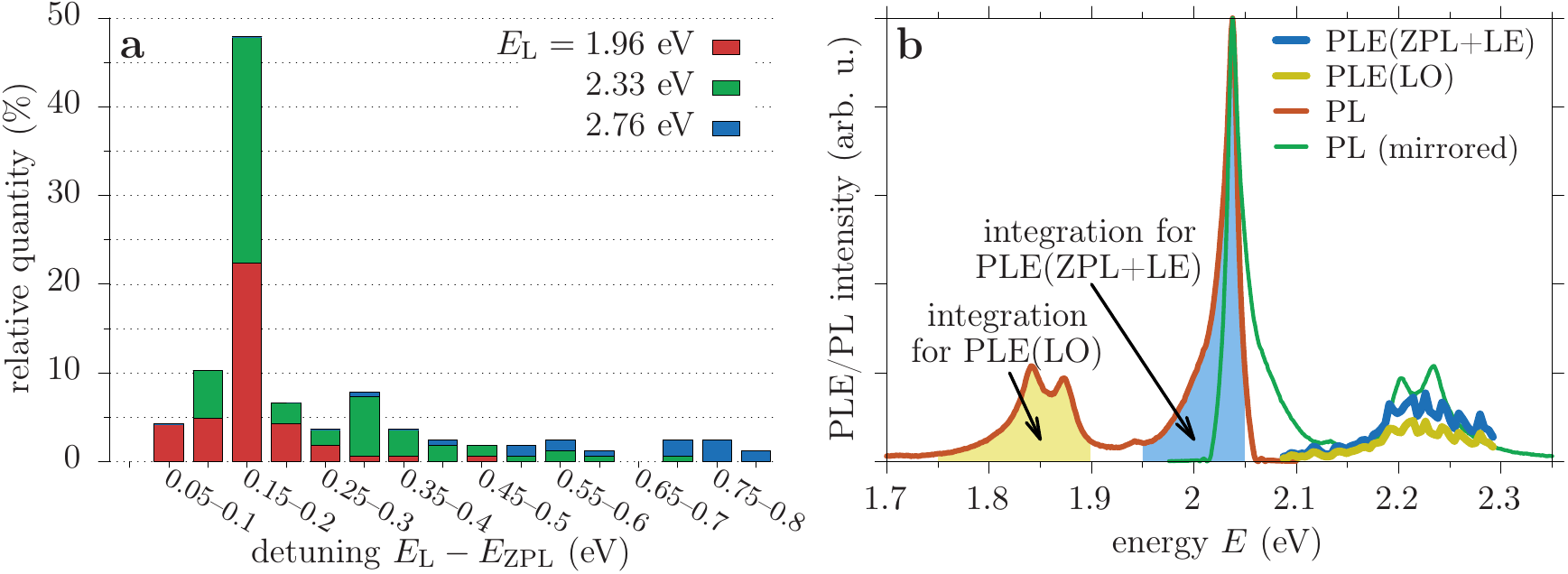}
\caption{{\bf LO phonon assisted absorption.} (a) Histogram of ZPL energies with respect to their detuning from the exciting laser. The colors mark the different laser excitations from Fig.~\ref{fig:2}(a). (b) Photoluminescence excitation (PLE) data retrieved from the ZPL (blue solid line) and the LO sideband (yellow solid line). The room temperature PL spectrum is plotted in orange with the shaded areas marking the integration areas to retrieve the PLE signals. The mirrored PL spectrum is given in green.}
\label{fig:7}
\end{figure}

We now focus again on the distribution of ZPL energies in Fig.~\ref{fig:2}(a). We find that many of the green dots cluster around $E_{\rm ZPL} = 2.15$~eV, while the red dots cluster around 1.8~eV. These energies are approximately 150--200~meV below the respective photon energy of the laser used for excitation (dashed lines), which is the range of LO phonon energies. This suggests that LO-phonon assisted absorption could efficiently drive the emitters. Therefore, we plot a histogram of detunings between the exciting laser energy $E_{\rm L}$ and the ZPL energy $E_{\rm ZPL}$ in Fig.~\ref{fig:7}(a), where the colors correspond to the colors in Fig.~\ref{fig:2}(a). In total, nearly 50\% of all 165 investigated emitters appear between 150~meV and 200~meV below the laser energy. The remaining emitters are almost equally distributed over the remaining 800~meV. This agrees with the findings in Ref.~\cite{jungwirth2017opt}.

To confirm that the emitters are efficiently excited by LO-phonon assisted absorption, we have performed PLE measurements at room temperature to obtain information about the absorption properties of the emitters at different photon energies. The results are shown in Fig.~\ref{fig:7}(b). We chose an emitter which has a strong LO phonon sideband in the PL spectrum, as shown with the orange line. At the high-energy side of the ZPL at 2.03~eV the spectrum is again cut off by a longpass filter. We measure PL spectra for a wide range of different excitation energies from 2.09~eV to 2.3~eV as shown in the Supplementary Material. After fitting and subtracting the background and Raman lines as described in the Supplementary Material, we extract the PLE data by integrating the PL spectra either over the ZPL, i.e., from 1.95~eV to 2.05~eV (blue shaded area), or over the LO phonon sideband, i.e., from 1.75~eV to 1.9~eV (yellow shaded area). The PLE data for the ZPL and for the LO phonon sideband are shown in blue and yellow, respectively. We find that both PLE data sets exhibit a pronounced maximum between 2.2~eV and 2.3~eV. This energy range nicely agrees with the phonon sidebands of the mirrored PL spectrum (green line). This observation demonstrates that the excitation of the emitter is very efficient via the LO-phonon assisted process.

Our study shows that the LO-phonon assisted excitation of the excited defect state provides an optimized way to drive the system. Therefore we can conclude that it is conceivable that we select a specific ZPL energy $\hbar\omega_0$, when pumping the system with energies between $\hbar\omega_0+160$~meV and $\hbar\omega_0+200$~meV, i.e, via a LO-phonon assisted absorption. This provides a strategy of isolating single-photon emitters with a desired emission energy from the wide range of possible transitions.

\section{Conclusions}

We have shown that the coupling to different phonon modes plays a crucial role for localized light emitters in hBN. On the one hand, coupling to bulk phonons leads to an asymmetric broadening of the ZPL in the case of LA modes and to the appearance of prominent sidebands well separated from the ZPL in the case of LO modes. On the other hand the coupling to a local mode oscillation contributes to the broadening of the ZPL at room temperature, while the respective side band is discernible at cryogenic temperatures. By fitting measured PL spectra with our theoretical model we were able to extract parameters for the wave function geometries of the emitters. We have shown that the distance between the positive and the negative charge center may be connected to the energy of the ZPL, supporting the assignment of the ZPL energy spread with the Stark effect caused by nearby charges. It was also possible to find reasonable values for the deformation potential of electrons and holes in hBN. By measuring PLE spectra, we have demonstrated that these sidebands are also present in absorption and lead to an efficient absorption via LO-phonon assisted transitions. Finally we have shown that it is possible to preferentially select emitters with a desired transition from the wide range of emission energies possible in hBN by aiming at the LO-phonon assisted absorption. Our results lead to a deeper understanding of the fundamental properties of color centers in hBN and pave the way to a tailored control of the excitation process.

\ack

We gratefully acknowledge support by the research group FOR 1493 of the Deutsche Forschungsgemeinschaft. We thank Michael Rohlfing for fruitful discussions.

\section*{References}
\providecommand{\newblock}{}

\pagebreak

\setcounter{page}{1}
\setcounter{section}{0}
\setcounter{figure}{0}
\setcounter{enumiv}{0}

\begin{center}
\Large{\bf Supplementary Material}
\end{center}

\section{Experimental details}

\subsection{Sample preparation}

\begin{figure}[h]
\centering
\includegraphics[width=0.85\columnwidth]{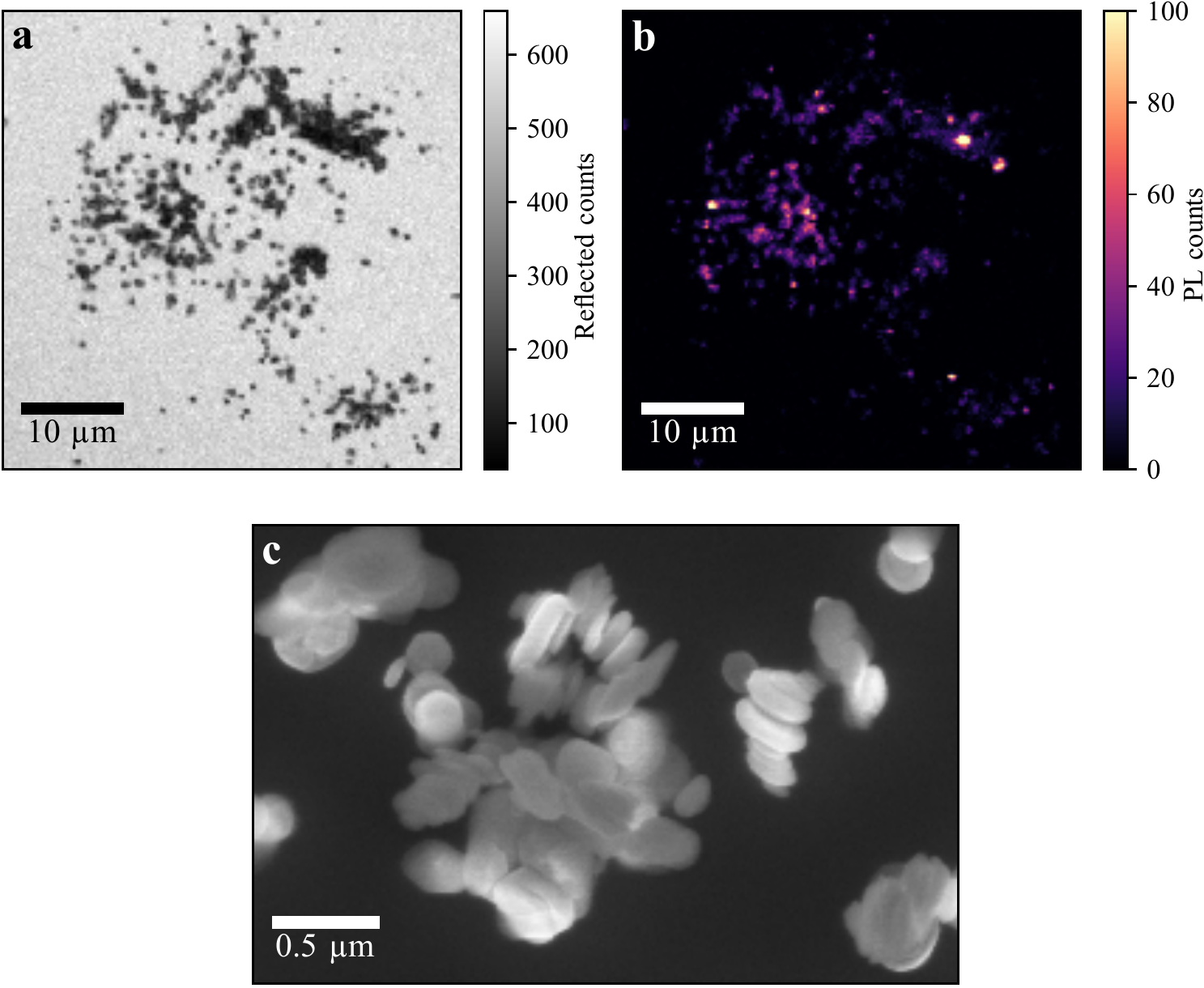}
\caption{(a) Reflectance and (b) photoluminescence (PL) image of hBN nano-powder exfoliated onto SiO2/Si substrate. (c) SEM image of hBN nano-powder. The shape of the hBN nano-powder ({\it Sigma Aldrich}) is disc-like with diameters ranging from tens to hundreds of nanometers. The thickness of the discs varies from several nanometers up to hundreds of nanometers.}
\label{fig:S1}
\end{figure}

Single-photon emitters (SPE) in hexagonal boron nitride are observed in hBN nano-powder ({\it Sigma-Aldrich}, grain size $<$150~nm). The powder is micromechanically exfoliated using scotch tape and stamped onto a Silicon substrate with a 80~nm or 270~nm thick thermal oxide layer on top. Figure~\ref{fig:S1} shows exemplary images of the nano powder sample. Panels (a) and (b) show a reflectance and PL image of the same sample area, respectively. The dark spots in (a) are clusters of the hBN powder crystals. The PL map in (b) shows bright spots at the positions of the clusters, where each spot stems from one or more defect centers. The SEM image in (c) shows the detailed stacked disc-like structure of the individual hBN crystals.

\subsection{Photoluminescence and photoluminescence excitation spectroscopy}
The photoluminescence experiments are carried out in a homebuilt confocal microscope. For excitation at 2.76~eV (450~nm) a diode laser and at 2.33~eV (532~nm) a frequency-doubled Nd:YAG laser is used. For recording the PLE spectra, the emitters are excited with a tunable laser light source based on a continuous-wave (cw) optical parametric oscillator (OPO). The turn-key system (from {\it H\"ubner Photonics}) covers the wavelength range of 450~nm -- 650~nm (2.76~eV -- 1.91~eV).

For the room-temperature measurements, the laser is focussed to the diffraction limit by an objective lens (numerical aperture NA=0.9), resulting in a focus size of $\approx 400$~nm. The excitation power is 100~\textmu W and kept constant for all excitation energies.

For the temperature-dependent measurements, the sample resides in a flow-cryostat cooled by liquid Helium. In this experiment, the laser is focussed on the sample using an objective lens with a NA of 0.75. In both cases, the same objective lens is used to collect the photoluminescence of the single emitters and the PL is analyzed in a spectrometer with an attached liquid nitrogen-cooled CCD camera. A longpass filter at 2.06~eV is used to remove stray light of the laser from the PL signal.

\section{Theory}

\subsection{Phenomenological model}

As explained in the main text we use a phenomenological fit function to extract relative spectral contributions from the phonon side bands of 165 different emitters.

The full fit function consists of three parts with
\begin{equation}
I(E) = I_{\rm ZPL}(E) + I_{\rm LO}(E) + I_{\rm LE}(E)\ ,
\end{equation}
where the zero phonon line (ZPL) is given by
\begin{equation}
I_{\rm ZPL}(E) =  \frac{A_{\rm ZPL}\gamma_{\rm ZPL}}{(E-E_{\rm ZPL})^2 + \gamma_{\rm ZPL}^2} \ .
\end{equation}
The longitudinal optical (LO) phonon side bands are assumed to have the same width as the ZPL, which leads to the function
\begin{eqnarray}
I_{\rm LO}(E) &=& I_{{\rm LO}_1}(E)+ I_{{\rm LO}_2}(E)\\
		&=& \frac{A_{{\rm LO}_1}\gamma_{\rm ZPL}}{[E-(E_{\rm ZPL} - E_{{\rm LO}_1} )]^2 + \gamma_{\rm ZPL}^2} \nonumber\\
		&+& \frac{A_{{\rm LO}_2}\gamma_{\rm ZPL}}{[E-(E_{\rm ZPL} - E_{{\rm LO}_2} )]^2 + \gamma_{\rm ZPL}^2}\ ,
\end{eqnarray}
where $E_{{\rm LO}_{1,2}}$ are the LO phonon energies.\\
Because the considered low energy (LE) phonon contributions have energies of $E_{{\rm LE}_1}=14$~meV and $E_{{\rm LE}_2}=30$~meV they are in the range of the thermal energy at room temperature $k_{\rm B}T\approx 25$~meV. Therefore we take phonon emission ($I_{{\rm LE}_j}(E)$) and absorption ($I_{{\rm LE}_j}^{\rm abs}(E)$) processes into account. Additionally we consider two phonon processes ($I_{{\rm LE}_j}^{2\rm ph}(E)$) for these modes and combination of LO and LE processes. Therefore, the LE phonon side bands read
\begin{equation}
I_{\rm LE}(E) = \sum_{j=1,2} \left( I_{{\rm LE}_j}(E) + I_{{\rm LE}_j}^{\rm abs}(E) + I_{{\rm LE}_j}^{2\rm ph}(E) \right)\ ,
\end{equation}
The single LE phonon lines read
\begin{eqnarray}
I_{{\rm LE}_j}(E) &=& \frac{A_{{\rm LE}_j}\gamma_{{\rm LE}_j}}{[E-(E_{\rm ZPL} - E_{{\rm LE}_j} )]^2 + \gamma_{{\rm LE}_j}^2}\ ,\\
I_{{\rm LE}_j}^{\rm abs}(E) &=& \frac{n_j}{n_j+1} \frac{A_{{\rm LE}_j}\gamma_{{\rm LE}_j}}{[E-(E_{\rm ZPL} + E_{{\rm LE}_j} )]^2 + \gamma_{{\rm LE}_j}^2}\ ,
\end{eqnarray}
where $n_j$ is the thermal occupation of LE mode $j$ from the Bose-Einstein distribution. to reduce the number of fit parameters we assume that the weights of the two phonon processes scale like the single LE phonon processes. This leads to the fit function
\begin{eqnarray}
I_{{\rm LE}_j}^{2\rm ph}(E) &=& \sum_{k=1,2}\Bigg\{ \frac{A_{{\rm LO}_k}}{A_{\rm ZPL}}\left[  I_{{\rm LE}_j}(E-E_{{\rm LO}_k}) + I_{{\rm LE}_j}^{\rm abs}(E-E_{{\rm LO}_k}) \right] \nonumber\\
&& \quad \ \ \, +  \frac{A_{{\rm LE}_k}}{A_{\rm ZPL}}\left[  I_{{\rm LE}_j}(E-E_{{\rm LE}_k}) + I_{{\rm LE}_j}^{\rm abs}(E-E_{{\rm LE}_k}) \right] \Bigg\}\ .
\end{eqnarray}

\subsection{Model parameters}
Motivated by Ref.~\cite{serrano2007vib}, for the LO phonon energies we consider two constant energies and for the LA phonons linear dispersions. We distinguish between the in-plane and the out-of-plane direction with
\begin{eqnarray*}
\hbar\omega_{{\rm LO}_1}(q_r) = E_{{\rm LO}_1}\ , \qquad & \hbar\omega_{{\rm LO}_2}(q_r) = E_{{\rm LO}_2}\ , &  \qquad \omega_{\rm LA}(q_r) = c_r q_r\\
\hbar\omega_{{\rm LO}_1}(q_z) = E_{{\rm LO}_1}\ ,  \qquad & \hbar\omega_{{\rm LO}_2}(q_z) = E_{{\rm LO}_1}\ , &  \qquad \omega_{\rm LA}(q_z) = c_z q_z
\end{eqnarray*}
The sound velocities are chosen to $c_z=3.44$~nm/ps~\cite{bosak2006ela} for the out-of-plane direction and for the in-plane direction we consider the mean value of the two high symmetry directions with $c_r=16$~nm/ps, which we extract from the dispersion relations in Ref.~\cite{michel2009the}.
Also the dielectric constants are different in the two distinct lattice directions with~\cite{geick1966nor}
\begin{eqnarray*}
\varepsilon_{\infty}(q_r) = 4.95\ , &\qquad & \varepsilon_{\rm s}(q_r) = 7.04\ ,\\
\varepsilon_{\infty}(q_z) = 4.10\ , &\qquad & \varepsilon_{\rm s}(q_z) = 5.09
\end{eqnarray*}
We interpolate between the two given directions via
\begin{eqnarray*}
\omega_{\rm j}(\bf q) &=& \cos^2(\varphi_q) \omega_{\rm j}(q_r) + \sin^2(\varphi_q)\omega_{\rm j}(q_z) \\
\varepsilon(\bf q) &=& \cos^2(\varphi_q) \varepsilon(q_r) + \sin^2(\varphi_q)\varepsilon(q_z)
\end{eqnarray*}
where $\varphi_q = {\rm atan}(q_z/q_r)$ is the angle of ${\bf q}$ with respect to the hBN-plane.

\newpage
\section{Temperature-dependent photoluminescence measurement}
\label{sec:temp}

\begin{figure}[h]
\centering
\includegraphics[width=\columnwidth]{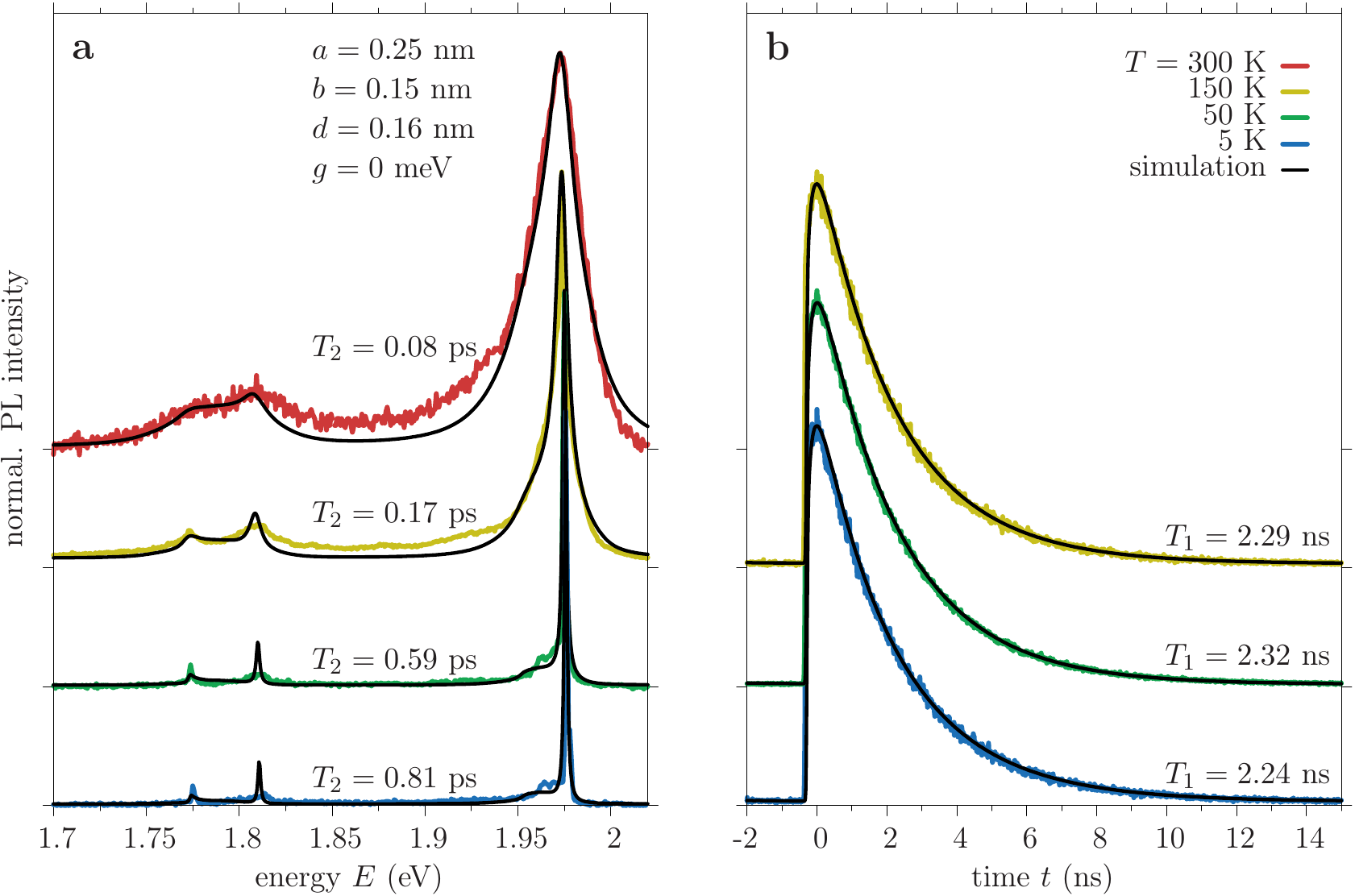}
\caption{Temperature-dependent PL spectra of one localized emitter. The temperature decreases from top to bottom (red to blue). The simulation is shown in green. The temperatures $T$ are given next to each spectrum.}
\label{fig:S2}
\end{figure}

To confirm the conclusions from the temperature-dependent measurements drawn in the main text, in Fig.~\ref{fig:S2} we show a second data set for a different localized emitter.

In Fig.~\ref{fig:S2}(a) the measured PL spectra (color) are shown for decreasing temperatures from top to bottom. The corresponding simulations are shown as black lines. The parameters used are given in the plot next to each curve. Here only the dephasing time $T_2$ is adjusted for each temperature, while the other parameters are kept constant. We did not find strong indications for a contribution of the local mode, leading to a coupling strength of $g=0$ in the simulations. The deformation potential couplings are the same as in the main text ($D_{\rm e}=0.4$~eV and $D_{\rm h}=-0.2$~eV). We again find that the ZPL strongly narrows when reducing the temperature.

In addition we measured the lifetime of the PL signal for different temperatures. These measurements are carried out in the same setup as the low-temperature PL measurements (see Sec. 1.2). However, the emitters are excited by a tunable femtosecond fiber laser system ($\approx 250$~fs pulse length, $40$~MHz repetition rate) at an energy of $2.16$~eV~\cite{sotier2009fem}. The photoluminescence is detected using a Picoquant PDM Series single-photon counting module (timing accuracy $50$~ps) and the PL decay measurement is performed with a Becker \& Hickl SPC-130-EM time-correlated single-photon counting card. The excitation power is kept constant for all measurements at $150$~\textmu W. The results are shown in Fig.~\ref{fig:S2}(b). We extract the lifetime $T_1$ by fitting the data with a convolution of the instrument response function and a single exponential decay (black curves). We find that the lifetime is almost constant $T_1\approx 2.3$~ns for all considered temperatures. These values are on the same time scale as those of defect centers in diamond~\cite{batalov2008tem}. 

In Refs.~\cite{dietrich2017nar,tran2017res} it was shown, that the homogenous linewidth of the emitters at low temperatures is in the sub-\textmu eV range and the corresponding $T_2$  in the ns range. In this case, the dephasing is dominated by the lifetime of the excited state, i.e., additional dephasing effects such as spectral wandering or pure dephasing vanish. This supports our observed trend of a strong narrowing of the ZPL with decreasing temperature. However, the values for low temperatures (5 K and 50 K) are underestimated because the measured line width is on the order of the limited resolution of the spectrometer (100 \textmu eV). Hence, we are not able to draw reliable conclusions about the low-temperature $T_2$ times from our data. At room temperature, we find that the radiative lifetime is four orders of magnitude longer than the dephasing time. Therefore, we can conclude that the dephasing due to the radiative lifetime is negligible compared to other dephasing mechanisms.

\newpage
\section{Color centers in diamond}

\begin{figure}[h]
\centering
\includegraphics[width=0.55\columnwidth]{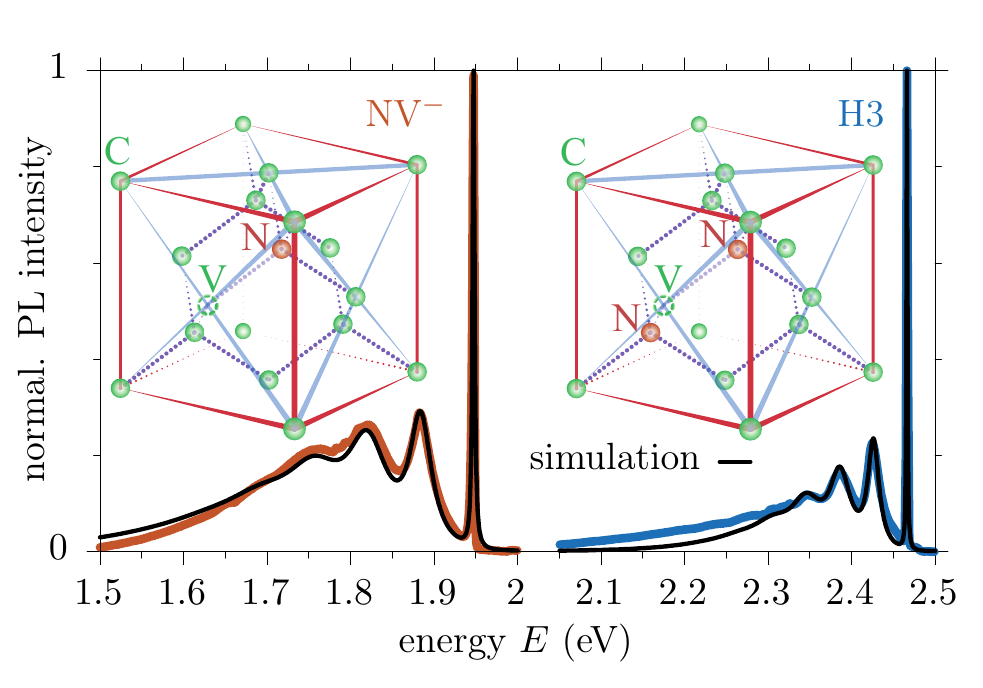}
\caption{PL spectra of color centers in diamond. Measured PL intensity for the NV$^-$ (orange) and the H3 defect (blue). Simulation in black. The insets show the atomic structures of the defects.}
\label{fig:S3}
\end{figure}

Figure~\ref{fig:S3} shows PL spectra of the nitrogen vacancy (NV$^-$) and the H3 color center in diamond in orange, respectively. The single-crystal diamond plate used for this experiment has a lateral size of $3\times 3$~mm$^2$ and a thickness of 1~mm and is produced by a high-pressure, high-temperature process ({\it element6}) and contains approximately 200~ppm nitrogen. For the excitation of the NV$^-$ center (H3-center) we used a 532~nm laser diode (405~nm laser diode) with a power of 100~\textmu W at the sample position. The focus diameter is 1.2~\textmu m for 532~nm excitation and 1.4~\textmu m for 405~nm excitation. Both spectra were measured at cryogenic temperatures. The blue curves represent the calculated spectra, where the coupling to a single local mode was considered. The model parameters with the best agreement with the experiment are listed in Tab.~\ref{tab:diamond}.

\begin{table}[h]
\caption{\label{tab:diamond}Model parameters for the simulations in Fig.~\ref{fig:S3}. The table shows the defect type, the ZPL energy $\hbar\omega_0$, the emitter dephasing time $T_2$ and the local mode energy $\hbar\omega_{\rm LOC}$, its strength $g$ and its width $\hbar\Delta$.}
\begin{tabular}{@{} c | c c c c c }
defect & $\hbar\omega_0$ (eV) & $T_2$ (ps) & $\hbar\omega_{\rm LOC}$ (meV) & $g/\omega_{\rm LOC}$ & $\hbar\Delta$ (meV)\\
\hline
NV$^-$ & 1.95 & 0.45 & 65.16 & 0.87 & 14.7 \\
H3 & 2.47 & 0.95 & 40.8  & 0.82 & 7.5 
\end{tabular}
\end{table}

\newpage
\section{PLE measurements}

\begin{figure}[h]
\centering
\includegraphics[width=0.49\textwidth]{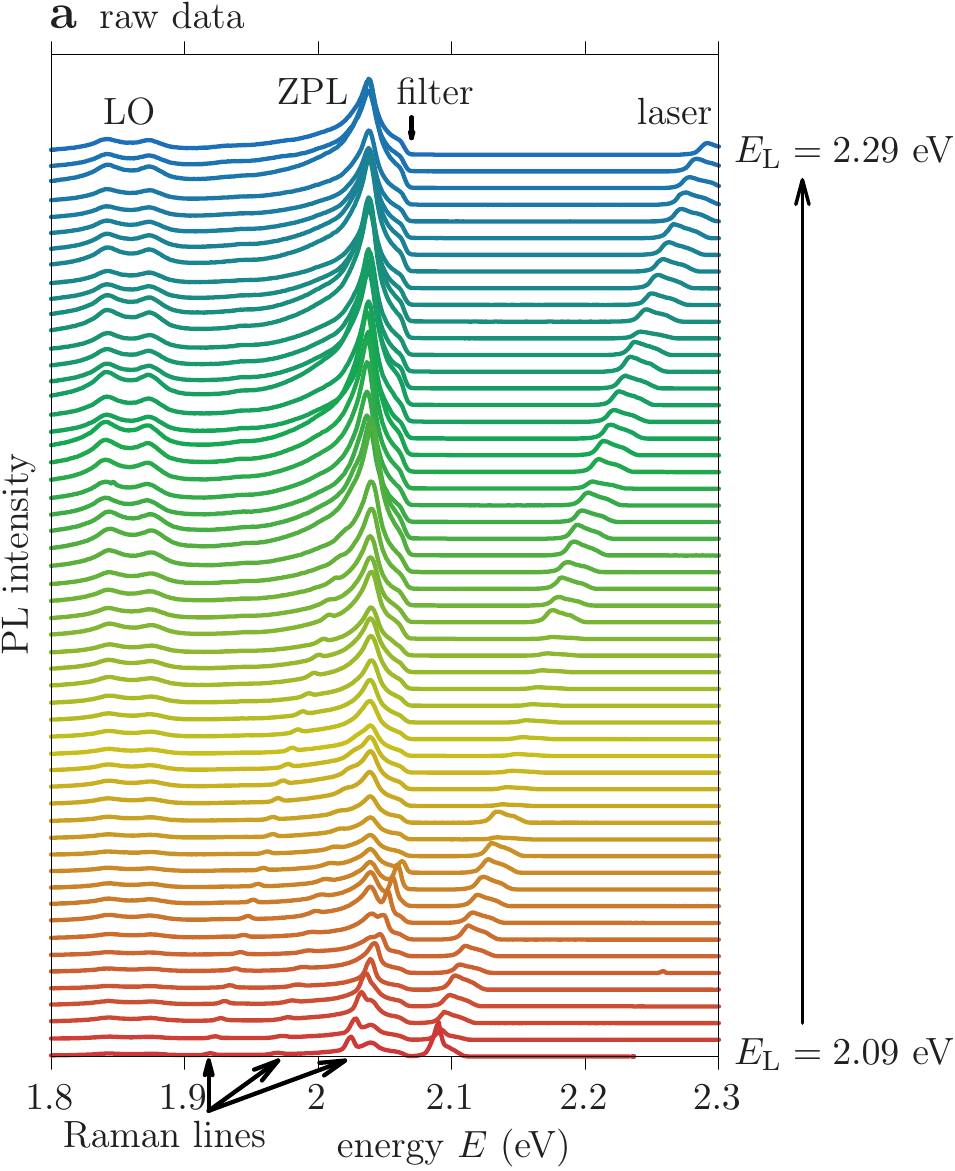}
\includegraphics[width=0.49\textwidth]{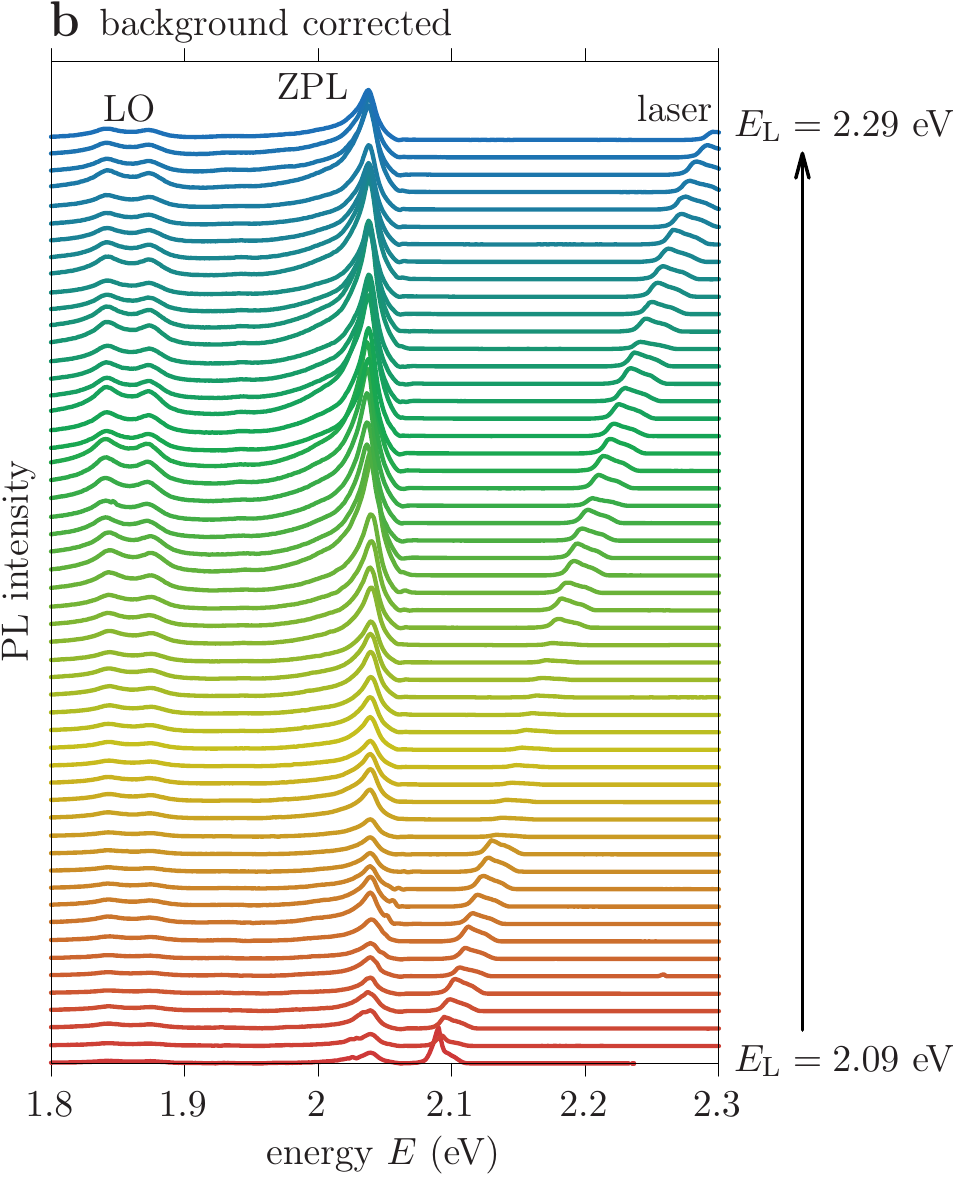}
\caption{PL spectra for different excitation energies. Low excitation energies at the bottom ($E_{\rm L}=2.09$~eV), high energies at the top ($E_{\rm L}=2.29$~eV) (blue to red). (a)~Raw PL data. (b) Background- and Raman-line-corrected spectra.}
\label{fig:S4}
\end{figure}

In Fig.~\ref{fig:S4}(a) PL spectra for different excitation energies are shown. The ZPL maxima and the phonon sidebands reside on a background signal. For excitation energies between 2.09~eV and 2.22~eV, Raman lines appear, which shift with the laser energy. The Raman shifts of the three lines are 63~meV, 118~meV, and 168~meV. We attributed the first two to the Si substrate~\cite{uchinokura1972ram} and the third one to the LO$_1$ energy~\cite{reich2005res}. To extract the intensity of the ZPL and the phonon sidebands, we fit the background by an exponential multiplied by an error function to reproduce the edge of the filter. The Raman lines are modeled by Gaussians. The background- and Raman-line-corrected spectra are shown in Fig.~\ref{fig:S4}(b).

For small excitation energies around 2.1~eV, which are very close to the ZPL energy, The PL spectra are very weak. Therefore both PLE signals in Fig.~7 in the main text drop significantly. This surprising observation indicates that in the energetic range where excitation processes - including the local mode - should occur, the emitter is not excited. Looking at the PL spectra in Fig.~\ref{fig:S4}(a) we see that for small excitation energies the ZPL overlaps with a strong Raman line, which might spoil the evaluation of the PLE data. Apart from that, the deviation between the PLE spectrum and the mirrored emission spectrum remains unclear and deserves additional investigation which is however beyond the scope of this paper. 

\newpage

\section*{References}
\providecommand{\newblock}{}

\end{document}